\documentclass[11pt,reqno]{article}
\usepackage{amsmath}
\usepackage{amsfonts}
\usepackage{amssymb}
\usepackage{latexsym}
\usepackage[dvips]{graphicx}
\usepackage{epsf}
\usepackage{color}

\usepackage{url}

\allowdisplaybreaks \numberwithin{equation}{section}

\newcommand{\U}{{\mathbf U}}
\newcommand{\myincludegraphics}[3]{\mkern10mu \raisebox{- #1 pt}{\scalebox{#2}{\includegraphics{#3}}}}

\newcommand{\Tr}{{\rm Tr}}
\newcommand{\tr}{{\rm tr}}

\title{On the non-local heat kernel expansion}

\author{Alessandro Codello$^1$ and Omar Zanusso$^2$\\
$^1$ \small{SISSA, Via Bonomea 265, I-34136 Trieste, Italy}\\
$^2$ \small{Institute of Physics, University of Mainz, Staudingerweg 7, D-55099 Mainz, Germany}\\
\small{\texttt{codello@sissa.it, zanusso@thep.physik.uni-mainz.de}}
}

\date{\,}

\begin{document}


\maketitle

\abstract{
We propose a novel derivation of the non-local heat kernel expansion,
first studied by Barvinsky, Vilkovisky and Avramidi, based on simple
diagrammatic equations satisfied by the heat kernel.
For Laplace-type differential operators we obtain the
explicit form of the non-local heat kernel form factors to second
order in the curvatures. Our method can be generalized easily to the derivation
of the non-local heat kernel expansion of a wide class of differential operators.
}


\section{Introduction}\label{Introduction}

The heat kernel plays a central role in many areas of mathematics
and physics. At present date two main expansion schemes have been developed
to calculate the heat kernel for general backgrounds. The first is
the early time or local expansion, which is also known as
the Schwinger-deWitt technique \cite{Vassilevich_2003,Barvinsky_Vilkovisky_1985,Vilkovisky_1992b}.
The second one is the non-local expansion studied by Barvinsky,
Vilkovisky and Avramidi \cite{Barvinsky_Vilkovisky_1987_1990,Avramidi_1990_2002}.
This latter case can be seen as an expansion in the curvatures.

In quantum field theory the local heat kernel expansion is usually
employed to compute ultraviolet divergences and anomalies, since these are
directly related to local heat kernel coefficients. The non-local
expansion is used instead to compute, in a fully covariant framework, the
finite part of the effective action \cite{Barvinsky_Vilkovisky_1987_1990,Avramidi_1990_2002}.
It also has many applications to the
effective average action \cite{Codello}
and to the spectral action \cite{Vassilevich_Spectral}.

The aim of the paper is to introduce a new diagrammatic technique
for the computation of the heat kernel, based on the combination
of a properly defined vertex-expansion and momentum-space rules.
These momentum-space rules can be derived from a single
functional that we call ``Laplacian action'', that serves as a device
for the computation of the heat kernel.
We will show how this method can be used to give a new and independent
derivation of the non-local heat kernel expansion first derived in
\cite{Barvinsky_Vilkovisky_1987_1990}.
We will consider the non-local expansion
for second order covariant Laplacians on a general boundaryless $d$-dimensional
manifold with arbitrary gauge connection.
A similar method has been proposed in \cite{Moss:1999wq},
but our presentation has the advantage of bringing
all the strength of Feynman-diagrams' reduction techniques
into the computation of the heat kernel expansion.

For a review of the more mathematical and geometrical aspects of the heat kernel see \cite{Rosenberg_1997},
while for a physicist's perspective see \cite{Mukanov_Winitzki_2007,Esposito_1998}.
A fully covariant scheme of the expansion of the heat kernel and some related operator may be found in \cite{Salcedo:2006pv}.

\section{Basic definitions}

Let us work on a general boundaryless $d$-dimensional manifold,
equipped with a Riemannian metric $g_{\mu\nu}$ and inverse $g^{\mu\nu}$.
We call the unique torsionless connection compatible with the metric $\Gamma_\mu$
and the associated covariant derivative $\nabla_\mu$, so that $\nabla_\mu g_{\nu\rho}=0$.
We also introduce a general vector bundle with connection $A_\mu$
and define the full covariant derivative to be $D_\mu = \partial_\mu + A_\mu$.
We do not specify further the structure of the bundle, it suffices to
know that the connection $A_\mu$ is expanded in terms of a basis of generators $T^a$
as $A_\mu=A_\mu^a T^a$, where repeated indices always imply
summation. Notice that the connection $A_\mu$ may coincide with the Levi-Civita
connection $\Gamma_\mu$ in some situations, for example when the vector bundle
is the tangent bundle itself.

The curvature of the connection is given by
\begin{eqnarray}\label{curvatures}
 \Omega_{\mu\nu} &\equiv& \left[D_\mu,D_\nu\right]\nonumber\\
 &=&
 \partial_\mu A_\nu-\partial_\nu A_\mu+\left[A_\mu,A_\nu\right]\, ,
\end{eqnarray}
and has components $F_{\mu\nu}=F_{\mu\nu}^a T^a$.
In the simple case in which \eqref{curvatures}
acts on the tangent vector bundle, we have by definition that $\Omega_{\mu\nu}={\cal R}_{\mu\nu}$,
where ${\cal R}_{\mu\nu}$ is the Riemann curvature $2$-form.
In components, the Riemann curvature $2$-form is the Riemann tensor $R_{\mu\nu}{}^\alpha{}_\beta$.
The Ricci tensor is defined as $R_{\mu\nu}\equiv R_{\alpha\mu}{}^\alpha{}_\nu$,
while the scalar curvature is defined as $R\equiv g^{\mu\nu}R_{\mu\nu}$.

We define a general Laplace-type operator to be a second order differential operator of the form
\begin{equation}\label{laplacian}
 \Delta \equiv -g^{\mu\nu}D_\mu D_\nu + \U
 = -D^2 + \U
 \,,
\end{equation}
where an endomorphism $\U$ over the vector bundle has been introduced.
From a physicist perspective, we may call $\U$ the potential term of the Laplace operator
and its components are defined as $\U=\U^a T^a$. For short, $\Delta$ will often be called simply Laplacian.

The heat kernel $K^{s}(x,y)$ is a bi-tensor of density weight $1/2$ defined through the following partial differential
equation and boundary condition:
\begin{equation}\label{HK_1}
 \left(\partial_{s}+\Delta_{x}\right)K^{s}(x,y)=0\qquad\qquad K^{0}(x,y)=\delta(x-y)\,.
\end{equation}
Notice that, when the Laplacian acts on tensors of different weight, also the coordinate representation changes.
We temporarily adopt two different symbols to stress this feature.
For any given tensor $\psi$ of weight $w=1/2$, we have that
\begin{equation}
 \Delta_{x,w=1/2}  \psi(x) = g^{1/4}(x) \Delta_{x,w=0} (g^{-1/4}(x) \psi(x))\, .
\end{equation}
The operators $\Delta_{x,w=1/2}$ and $\Delta_{x,w=0}$ can be understood as the components of the same abstract operator $\Delta$,
when acting on tensors with different weight. An important property is that the two share the spectrum.
We could have, in principle, adopted a different weight normalization for \eqref{HK_1},
but the one we choose turns out to be particularly convenient in the following \cite{Mukanov_Winitzki_2007}.
The heat kernel equation \eqref{HK_1} can be interpreted as describing a continuous diffusion process on the manifold.
In this picture, the diffusing particles are all concentrated at the origin $x=y$, when the heat
kernel proper-time parameter $s$ is zero.
The heat kernel proper-time $s$ is related to the real time $t$ and the diffusion constant ${\cal D}$
through the relation $s = {\cal D} t$.

Equation \eqref{HK_1} is formally solved as
\begin{equation}\label{HK_2}
 K^{s}(x,y)=e^{-s\Delta_{x}}\delta(x-y)\, ,
\end{equation}
so that its trace is formally the following:
\begin{equation}\label{tr_HK_1}
 \Tr \, K^{s} = \left.\Tr \, e^{-s\Delta_x} \delta(x-y)\right|_{x=y} \equiv \Tr \, e^{-s\Delta} \, .
\end{equation}
In order to give a precise meaning to the rather formal expressions \eqref{HK_2} and \eqref{tr_HK_1},
we want to introduce a basis of orthonormal eigenfunctions
for $\Delta$.
Special care has to be taken when dealing with a non-flat
background where the initial condition in \eqref{HK_1} requires a
particular choice of the eigenfunction basis.
We introduce a basis $\phi_n$ of eigenmodes of weight $w=1/2$ for the Laplacian
\begin{equation}\label{HK_2.13}
 \Delta_{x}\phi_{n}(x)=\lambda_{n}\phi_{n}(x)\,,
\end{equation}
that is normalized as
\begin{equation}\label{basis2}
 \int d^{d}x\,
 \phi_{n}(x)\phi_{m}(x)=\delta_{mn}
 \qquad\qquad
 \sum_{n}\phi_{n}(x)\phi_{n}(y)=
 \delta(x-y)\, .
\end{equation}

Equation \eqref{HK_2} is written rigorously using the basis $\phi_n$ as
\begin{equation}\label{HK_3}
 K^{s}(x,y)=\sum_{n}e^{-s\lambda_{n}}\phi_{n}(x)\phi_{n}(y)\, ,
\end{equation}
where the initial condition of \eqref{HK_1} is satisfied thanks to \eqref{basis2}
\begin{equation}
K^{0}(x,y)=\sum_{n}\phi_{n}(x)\phi_{n}(y)=\delta(x-y)\,.
\end{equation}
Tracing \eqref{HK_3} and using the orthogonality relations in
\eqref{basis2}, we obtain
\begin{equation}\label{HK_4}
 \Tr K^{s}={\rm tr}\!\int d^{d}x\, K^{s}(x,x)=\sum_{n}e^{-s\lambda_{n}}\int d^{d}x\,\phi_{n}(x)\phi_{n}(x)=\sum_{n}e^{-s\lambda_{n}}\,.
\end{equation}
Note that in \eqref{basis2} there is no factor $\sqrt{g}$
in the integrals because of its weight normalization.
This will turn out to be particularly useful later on,
because it decouples the normalization of the basis from
the metric dependence of the heat kernel \eqref{HK_2}.

As a final remark of the section, we want to stress
one of the many useful applications of the heat kernel.
It is used in theoretical physics to give a precise meaning to and to evaluate functional traces.
In fact, every trace of an arbitrary positive function of the Laplacian $h(\Delta)$
is related to the heat kernel by means of a Laplace-transform
\begin{equation}\label{any_trace}
 \Tr h(\Delta) = \int_{0}^{\infty}ds\,\tilde{h}(s)\,\Tr e^{-s\Delta} \, ,
\end{equation}
where $\tilde{h}(s)$ is the inverse-Laplace transform of $h(x)$.
Therefore, in order to compute such a functional trace,
one simply needs to know the expansion of the trace of the heat kernel to any desired accuracy.
The local and non-local expansion schemes mentioned in section \ref{Introduction}
represent the two main expansion schemes with which \eqref{any_trace} has been
successfully applied to quantum field theory computations.

\section{Perturbative expansion of the heat kernel}

In this section we develop the perturbative expansion for the computation of the
heat kernel, where the covariant Laplacian is decomposed as the sum
of a non-interacting Laplacian $-\partial^{2}$ and of an
interaction $V$ in the following way:
\begin{equation}\label{HK_PT_1}
 \Delta\equiv-\partial^{2}+V\,.
\end{equation}
The potential $V$ contains $\U$, all terms proportional to the gauge
connection $A_\mu$ and all terms obtained by expanding the metric $g_{\mu\nu}=\delta_{\mu\nu}+h_{\mu\nu}$
in powers of the fluctuation $h_{\mu\nu}$.

We explicitly give two simple examples of the decomposition \eqref{HK_PT_1}.
The flat space limit is
\begin{equation}
 \Delta=-\partial^{2}+\U
\end{equation}
where simply $V=\U$;
the gauge Laplacian where $V$ contains all terms that vanish for $A_{\mu}=0$,
\begin{equation}
 \Delta
 =-\left(\partial_{\mu}+A_{\mu}\right)\left(\partial^{\mu}+A^{\mu}\right)=-\partial^{2}\underbrace{-2A_{\mu}\partial^{\mu}-\partial_{\mu}A^{\mu}-A_{\mu}A^{\mu}}_{V}\,.
\end{equation}
In both cases there is no distinction between different density weights.

We derive now the perturbative expansion in $V$ of the heat kernel.
First we need to compute the heat kernel
$K_{0,xy}^{s}$ of the operator $-\partial^{2}$, around which we
will perform the perturbative expansion.
Here and in the following we use the compact notation $K_{xy}^{s}\equiv K^{s}(x,y)$
and $\delta_{xy}\equiv\delta^{(d)}(x-y)$. We also define $\int_{x}\equiv\int d^{d}x$
and $\int_{q}\equiv\int\frac{d^{d}q}{(2\pi)^{d}}$.
Analogously to \eqref{HK_1},
we require that it satisfies the following equation with boundary condition:
\begin{equation}\label{HK_PT_2}
 \left(\partial_{s}-\partial_{x}^{2}\right)K_{0,xy}^{s}=0\qquad\qquad K_{0,xy}^{0}=\delta_{xy}\,.
\end{equation}
The differential equation \eqref{HK_PT_2} is easily solved in momentum space.
If we Fourier transform the heat kernel
\begin{equation}\label{HK_PT_3}
 K_{0,xy}^{s}=\int_{qq'}K_{0,qq'}^{s}\, e^{-i(xq+yq')}\,,
\end{equation}
then \eqref{HK_PT_2} becomes simply
\begin{equation}\label{HK_PT_4}
 \left(\partial_{s}+q^{2}\right)K_{0,qq'}^{s}=0\qquad\qquad K_{0,qq'}^{0}=\delta_{q+q'}\,.
\end{equation}
The solution of \eqref{HK_PT_4} is trivially seen to be
\begin{equation}
K_{0,qq'}^{s}=\delta_{q+q'}e^{-sq^{2}}\,.
\end{equation}
Transforming this solution back to coordinate space, completing the
square and performing the Gaussian integral
gives the following result:
\begin{equation}\label{HK_PT_5}
 K_{0,xy}^{s}=\frac{1}{(4\pi s)^{d/2}}e^{-\frac{(x-y)^{2}}{4s}}\,.
\end{equation}
This is the fundamental solution around which
we will construct the perturbative expansion.

To begin with, we first note that the heat kernel
\eqref{HK_PT_5} satisfies the composition rule
\begin{equation}\label{HK_PT_5.1}
 K_{0,xy}^{s_{1}+s_{2}}=\int_{z}K_{0,xz}^{s_{1}}K_{0,zy}^{s_{2}}\,.
\end{equation}
We define the operator $U_{xy}^{s}\equiv\int_{z}K_{0,xz}^{-s}K_{zy}^{s}$.
Using \eqref{HK_1} and \eqref{HK_PT_4}, we find it satisfies
\begin{eqnarray}\label{HK_PT_6}
 \partial_{s}U_{xy}^{s} & = & \int_{z}\left[\partial_{s}K_{0,xz}^{-s}K_{zy}^{s}+K_{0,xz}^{-s}\partial_{s}K_{zy}^{s}\right]\nonumber \\
 & = & \int_{z}\left[K_{0,xz}^{-s}(-\partial_{z}^{2})K_{zy}^{s}-K_{0,xz}^{-s}\Delta_{z}K_{zy}^{s}\right]\nonumber \\
 & = & -\int_{z}K_{0,xz}^{-s}V_{z}K_{zy}^{s}\nonumber \\
 & = & -\int_{zw}K_{0,xz}^{-s}V_{z}K_{0,zw}^{s}U_{wy}^{s}\,.
\end{eqnarray}
We use Dyson's series to solve equation \eqref{HK_PT_6} in a compact form
\begin{equation}\label{HK_U}
 U_{xy}^{s}=T\exp\left\{ -\int_{0}^{s}dt\,\int_{z}K_{0,xz}^{-t}V_{z}K_{0,zy}^{t}\right\} \, ,
\end{equation}
where the exponential is time-ordered with respect to the parameter $s$.
Substituting \eqref{HK_U} into \eqref{HK_PT_6} and using \eqref{HK_PT_5.1}, we immediately find that
\begin{equation}\label{HK_PT_7}
 K_{xy}^{s}=\int_{z}K_{0,xz}^{s}\, T\exp\left\{ -\int_{0}^{s}dt\,\int_{w}K_{0,zw}^{-t}V_{w}K_{0,wy}^{t}\right\} \,.
\end{equation}
Rescaling the integration variable in \eqref{HK_PT_7} as $t\rightarrow t/s$
and using \eqref{HK_PT_5.1}, we obtain the final formula for the perturbative
expansion of the heat kernel
\begin{eqnarray}\label{HK_PT_8}
 K_{xy}^{s} & = &
 \int_{z}K_{0,xz}^{s}\, T\exp\left\{ -s\int_{0}^{1}dt\int_{w}K_{0,zw}^{-st}V_{w}K_{0,wy}^{st}\right\}\nonumber \\
 & = & K_{0,xy}^{s}-s\int_{0}^{1}dt\,\int_{z}K_{0,xz}^{s(1-t)}V_{z}\, K_{0,zy}^{st}+\nonumber \\
 &  & +s^{2}\int_{0}^{1}dt_{1}\int_{0}^{t_{1}}\, dt_{2}\,\int_{zw}K_{0,xz}^{s(1-t_{1})}V_{z}\, K_{0,zw}^{s(t_{1}-t_{2})}V_{w}\, K_{0,wy}^{st_{2}}+O(V^{3})\,,
\end{eqnarray}
that represents the core of the covariant perturbation theory of the heat kernel \cite{Barvinsky_Vilkovisky_1987_1990}.
The finiteness of each term in this expansion is guaranteed if $V$ is smooth and the covergence has been extensively studied in \cite{Barvinsky_Vilkovisky_1987_1990}.
The terms are finite even if $V$ has a discontinuity of a derivative or a delta-function
singularity \cite{Bordag:2004rx}. As long as $\U, A_\mu$ are such that $V$ has the above properties and $g_{\mu\nu}$ is smooth, the integrals in \eqref{HK_PT_8} are convergent.

The expansion \eqref{HK_PT_8} is conveniently represented graphically
\begin{eqnarray}\label{HK_PT_8.1}
 K^s=
 \myincludegraphics{0}{0.30}{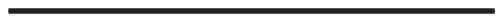}
 -s
 \myincludegraphics{2}{0.30}{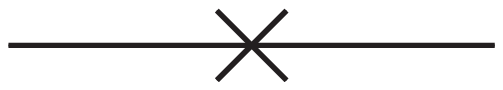}
 +s^2
 \myincludegraphics{2}{0.30}{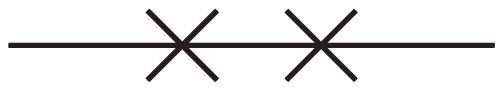}
 +\dots
\end{eqnarray}
where a continuous line represents a $K_{0,xy}$ factor,
while with each cross is associated an insertion of the interaction potential $V_{x}$.
The general term of the expansion \eqref{HK_PT_8} reads
\begin{equation}\label{HK_PT_8.2}{}
 (-s)^{n}\int_{0}^{1}dt_{1}\cdots\int_{0}^{t_{n-1}}\, dt_{n}\,\int_{z_{1}\cdots z_{n}}
 K_{0,x z_{1}}^{s(1-t_{1})}V_{z_{1}}K_{0,z_{1}z_{2}}^{s(t_{1}-t_{2})}\cdots
 V_{z_{n-1}}K_{0,z_{n}y}^{s t_{n}}
 \,.
\end{equation}
We now use the expansion \eqref{HK_PT_8} to derive the perturbative
expansion for the trace of the heat kernel. To do this we simply trace
\eqref{HK_PT_8}. This gives, in graphical form, the following
expansion for the heat kernel trace:
\begin{eqnarray}\label{HK_PT_9.1}
 {\rm Tr } \, K^s=
 \myincludegraphics{14}{0.30}{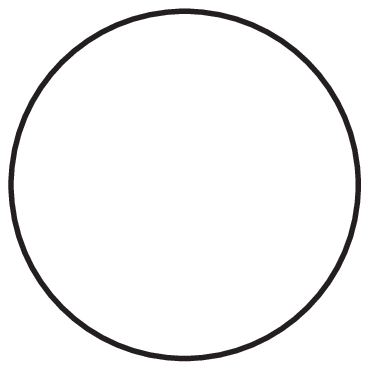}
 -s
 \myincludegraphics{15}{0.30}{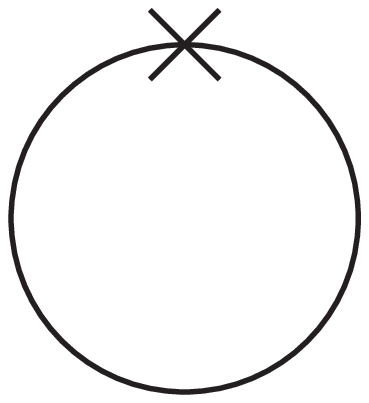}
 + s^2
 \myincludegraphics{17}{0.30}{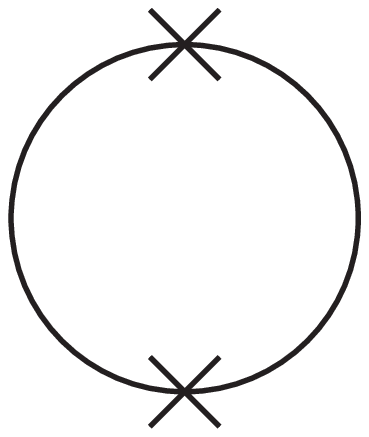}
 +\dots
\end{eqnarray}
and the general term in this expansion is now of the form
\begin{equation}\label{HK_PT_9}{}
 (-s)^{n}\tr\int_{0}^{1}dt_{1}\cdots\int_{0}^{t_{n-1}}\, dt_{n}\,\int_{z_{1}\cdots z_{n}}
 K_{0,z_{n}z_{1}}^{s(1-t_{1}+t_{n})}V_{z_{1}}K_{0,z_{1}z_{2}}^{s(t_{1}-t_{2})}\cdots
 K_{0,z_{n-2}z_{n-1}}^{s(t_{n-2}-t_{n-1})}
 V_{z_{n-1}}
 \,.
\end{equation}
Note that in \eqref{HK_PT_9} we used the cyclicity of the trace to
combine last and first flat heat kernels of \eqref{HK_PT_8}.

Let $\Phi$ denote collectively the set of fields $\Phi\equiv\{\U,A_\mu,h_{\mu\nu}\}$.
It should be clear that both the trace of the heat kernel \eqref{HK_PT_9.1} and the heat kernel \eqref{HK_PT_8.1}
are functionals of $\Phi$.
We define a $n$-point function of the trace of the heat kernel \eqref{HK_PT_9.1} to be
\begin{equation}
  \frac{\delta^n \Tr K^s}{\delta \Phi(x_1)\dots \delta\Phi(x_n)}\,,
\end{equation}
while a similar definition can be adopted for the heat kernel as well.

To make profit of the
perturbative expansions \eqref{HK_PT_8.1} and \eqref{HK_PT_9.1} we derived for the heat kernel and its trace,
we need to devise a method to actually compute the various contributions.
The strategy that we propose in this paper is simple but effective:
derive equations for the $n$-point functions of the heat kernel
by taking functional derivatives of the expansion \eqref{HK_PT_8}
with respect to $\mathbf{U},A_{\mu},h_{\mu\nu}$ and then set them to zero.
We then extract all possible information from each equation so derived,
by considering their Fourier transform in momentum space.
The equations for the $n$-point functions of the heat kernel trace
read as follows:
\begin{eqnarray}\label{diagrams}
 \left.\Tr K^s\right|_{\Phi=0}
 &=&
 \myincludegraphics{14}{0.30}{images/hk_loop.eps}\nonumber
 \\
 \left.\frac{\delta\Tr K^s}{\delta\Phi(x)}\right|_{\Phi=0}
 &=&
 -s
 \myincludegraphics{14}{0.30}{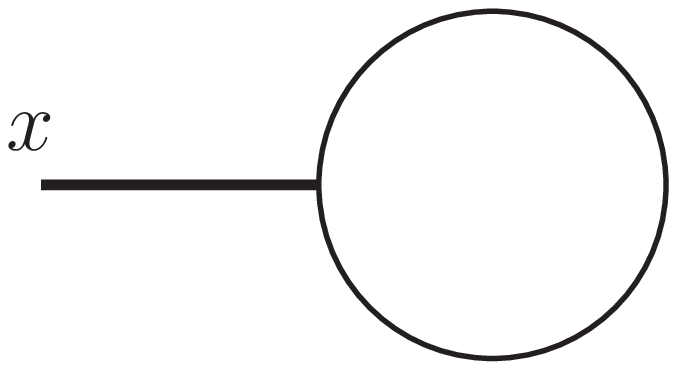}\nonumber
 \\
 \left.\frac{\delta^2\Tr K^s}{\delta\Phi(x)\delta\Phi(y)}\right|_{\Phi=0}
 &=&
 2 s^2
 \myincludegraphics{14}{0.30}{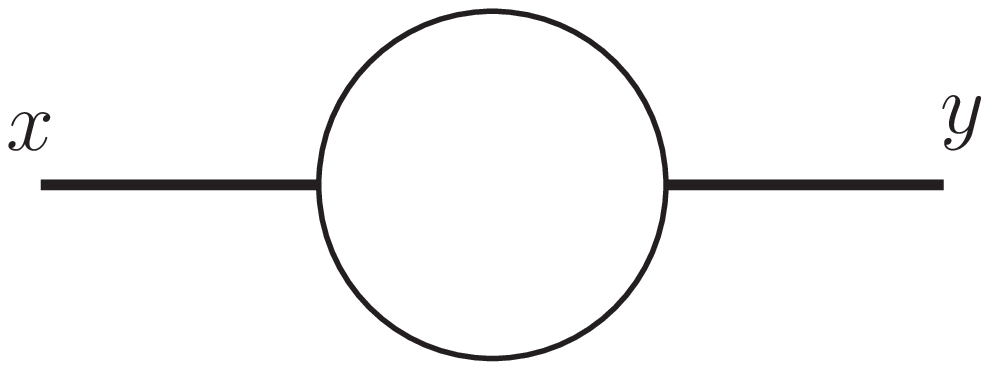}
 - s
 \myincludegraphics{18}{0.30}{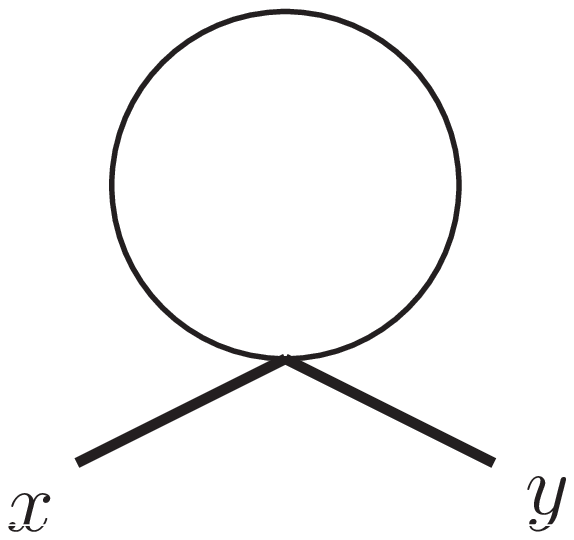}
\end{eqnarray}
and so on. These relations will be used, as we show
in the next section, to compute the heat kernel expansion order
by order in the curvatures.
The term curvatures is used in a generalized sense.
We include in the set of curvatures $\U$, $\Omega_{\mu\nu}$ and all tensors constructed with $R_{\mu\nu}{}^\alpha{}_\beta$.
In particular we will use the equation
for the two-point function to determine the non-local form factors
at second order in the curvatures.

Our aim now is to obtain the momentum space rules to compute the various diagrams
present in the equations for $n$-point functions of the heat kernel
expansion \eqref{diagrams}. It is useful, first, to define a functional, that we will call
``Laplacian action'' and that serves as a simple device to compute
the vertices appearing in the expansion as Feynman rules.
We first introduce an auxiliary field $\phi$ taking values in the internal space and having density weight $w=0$.
The Laplacian action $L[\phi,\Phi]$ is defined to be a quadratic action in $\phi$,
whose Hessian is the Laplacian \eqref{laplacian} with the correct density weight.
This notion carries over straightforwardly to the case of more general differential operators of non-Lapalcian type.
The propagation of the field $\phi$ is directly related to the propagation of the heat kernel degrees of freedom.
By construction, the second derivative of $L[\phi,\Phi]$ with respect to $\phi$ is
\begin{equation}\label{HK_PT_10}
L^{(2;0)}[\phi;\Phi]_{xy}=g(x)^{1/4}\Delta[\Phi]_{xy}\,g(y)^{-1/4}\,,
\end{equation}
where $\Delta[\Phi]$ is the Laplacian appearing in \eqref{laplacian}.
The dependence of the Laplacian on $\Phi$ was made explicit for a better understanding.
In \eqref{HK_PT_10} two powers of the determinant of the metric are present,
so that $\Delta[\Phi]_{xy}$ are the components of the Laplacian when acting on tensors of density weight $w=0$
as we are correctly densitizing the formula to match with the basis \eqref{basis2}
we adopted.
Therefore, from \eqref{HK_PT_10} we obtain
\begin{eqnarray}\label{HK_PT_11}
 L[\phi;\Phi]
 &=&
 \frac{1}{2}\int d^{d}x \,g^{1/4}\,\phi\,\Delta[\Phi](\phi\,g^{-1/4})\nonumber\\
 &=&
 \frac{1}{2}\int d^{d}x \sqrt{g} \Bigl[ g^{\mu\nu} D_{\mu}(\phi\,g^{-1/4})D_{\nu}(\phi\,g^{-1/4}) + g^{-1/2}\U\phi^2\Bigr]
 \,.
\end{eqnarray}
We expand \eqref{HK_PT_10} in a functional Taylor series in $\Phi$,
\begin{equation}\label{HK_PT_12}{}
 L_{xy}^{(2;0)}[\phi;\Phi]
 =L_{xy}^{(2;0)}[0;0]+\int_{z}L_{xyz}^{(2;1)}[0;0]\Phi_{z}+\frac{1}{2}\int_{zw}L_{xyzw}^{(2;2)}[0;0]\Phi_{z}\Phi_{w}+O(\Phi^{3})
\end{equation}
to observe that $L_{xy}^{(2;0)}[0;0]=-\partial_{x}^{2}\delta_{xy}$
is precisely the flat space Laplacian around which we are expanding.
We can thus write the interaction potential defined in \eqref{HK_PT_1} as
\begin{equation}\label{HK_PT_13}
 V_{x}\delta_{xy}=\int_{z}L_{xyz}^{(2;1)}[0;0]\Phi_{z}+\frac{1}{2}\int_{zw}L_{xyzw}^{(2;2)}[0;0]\Phi_{z}\Phi_{w}+O(\Phi^{3})\,.
\end{equation}
This shows that the vertices in the diagrams \eqref{diagrams} are precisely those constructed
from the Laplacian action \eqref{HK_PT_11}.
The Laplacian action generates all the interactions needed for the perturbative expansion of the heat kernel
and therefore has the role of the generating function of our method \cite{Vasiliev:1998cq}.

We can now state the Feynman rules for the computations of \eqref{diagrams} in momentum space.
Bolder lines represent external $\Phi$ lines, that are associated with functional derivatives w.r.t. $\Phi$ in \eqref{diagrams}.
Lighter lines instead represent the propagation of the auxiliary field $\phi$, that has the flat space heat kernel as propagator.
The arrows point in the direction of the momentum (when giving a vertex we always adopt the convention that all momenta are incoming).
The composition of propagators is derived from \eqref{HK_PT_8} for any number of insertions by taking the Fourier transform of \eqref{HK_PT_8.2} and chopping the $V$ insertions.
It is represented as
\begin{eqnarray}\label{general_propagator}{}
 \myincludegraphics{17}{0.60}{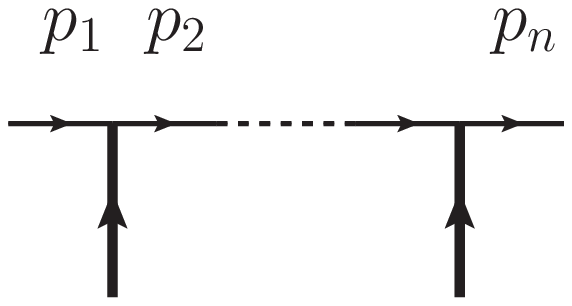}
 & = & \int_{0}^{1}dt_{1}\int_{0}^{t_{1}}\, dt_{2}...\int_{0}^{t_{n-2}}\, dt_{n-1}\, e^{-s(1-t_{1})p_{1}^{2}}e^{-s(t_{1}-t_{2})p_{2}^{2}}\cdots\nonumber\\
 &  & \cdots\, e^{-s(t_{n-2}-t_{n-1})p_{n-1}^{2}}e^{-st_{n-1}p_{n}^{2}}\,,
\end{eqnarray}
where bold lines are placed in correspondence of any chopped $V$ insertion for convenience.
The $V$ insertions, instead, appear as vertices. These are obtained by taking functional derivatives of \eqref{HK_PT_11}, evaluating the result in momentum space.
They are given by
\begin{eqnarray}\label{general_vertex}
 \myincludegraphics{30}{0.60}{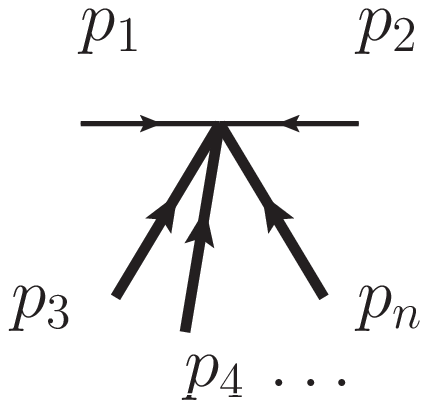}
 &=&
 L^{(2,n-2)}[0,0]_{p_1,p_2,\dots,p_n}
\end{eqnarray}
An integration over closed loop momenta has to be performed.
In the next section we show how to apply these diagrammatic rules to the calculation
of the heat kernel trace.

The existence of Feynman rules for the diagrams \eqref{general_propagator} and \eqref{general_vertex}
requires the presence of asymptotic momentum states in the manifold under consideration.
This, in turn, implies that the manifold has to be asymptotically $\mathbb{R}^d$.
In a more general situation, it is still possible to develop the construction of the expansion of the heat kernel \eqref{HK_PT_8}
around an operator that is more general than $-\partial^2$,
once a full basis of eigenstates of the latter is known.
This general treatment however would go beyond the scope of this work.

\section{Calculating the heat kernel trace}\label{section2}

In this section we give the computation of all the possible graphs appearing in \eqref{diagrams} using the rules \eqref{general_propagator} and \eqref{general_vertex}.
The results have to be compared with an appropriate ansatz for the curvature expansion of the heat kernel.
The most general expansion of the trace of the heat kernel \eqref{HK_2} up to second order in the curvatures
$\U$, $\Omega_{\mu\nu}$ and all tensors constructed from $R_{\mu\nu}{}^\alpha{}_\beta$ is \cite{Barvinsky_Vilkovisky_1987_1990}
\begin{eqnarray}\label{ansatz1}{}
 \Tr K^{s}
 & = &
 \frac{1}{(4\pi s)^{d/2}}\int d^{d}x\sqrt{g}\,\tr\Bigr\{
 g_0 \,\mathbf{1}
 +s \, g_{U,0} \, \U
 +s \, g_{R,0} \, \mathbf{1} R
 +s^{2}\Bigr[\mathbf{1}R_{\mu\nu}f_{Ric}(s\Box)R^{\mu\nu}
 \nonumber \\
 {}
 &  &
 +\mathbf{1}R\, f_{R}(s\Box)R
 +R\, f_{RU}(s\Box)\U
 +\U f_{U}(s\Box)\U
 +\Omega_{\mu\nu}f_{\Omega}(s\Box)\Omega^{\mu\nu}\Bigl]
 +{\cal O}(\mathcal{R}^{3})\Bigl\} \,.
\end{eqnarray}
where $\tr$ stands for the trace over the internal vector bundle and $\Box=-D^2$ is the Laplacian without the endomorphism term.
We refer the reader to appendix \ref{appendixA} for some mathematical details used in the contruction of \eqref{ansatz1}.
The expansion \eqref{ansatz1} has the name of covariant perturbation theory of the heat kernel and its validity obeys the condition
\begin{equation}\label{cpt_condition}
 \nabla\nabla{\cal R}\gg {\cal R}^2\,,
\end{equation}
where ${\cal R}$ is any element of the set $\left\{R_{\mu\nu},R,\U,\Omega_{\mu\nu}\right\}$ \cite{Barvinsky_Vilkovisky_1987_1990}.
We complement the condition \eqref{cpt_condition} by requiring that the manifold under consideration is asymptotically flat,
in order for momentum space rules \eqref{general_propagator} and \eqref{general_vertex} to exist.

We realize that there are three constants to be computed, $g_0$, $g_{U,0}$ and $g_{R,0}$.
Further, at quadratic order in the curvatures, five so-called  form factors $f_{Ric}(x)$, $f_{R}(x)$, $f_{RU}(x)$, $f_{U}(x)$ and $f_{\Omega}(x)$ appear \cite{Barvinsky_Vilkovisky_1987_1990}.
These represent an important departure from the standard Seeley-deWitt method to compute the heat kernel in local form (based on an expansion in $s$),
although it is possible to understand them as the resummation of infinite number of terms of the latter \cite{Avramidi_1990_2002}.
Each order in the curvatures of \eqref{ansatz1} has been normalized by an appropriate power of $s$, so that all the constants and all the form factors are dimensionless
if a canonical dimension is given to the covariant derivatives. It follows naturally, then, that the functions have to have the dimensionless product $s\Box$ as argument.

Notice that, since our manifold of choice is boundaryless all terms of the form $ \Box^n R$ and $\Box^n \U$ for $ n\ge 1$ are not present in \eqref{ansatz1}.
We will come back to these terms in appendix \ref{appendixB}.

\subsection{Endomorphism contribution to the heat kernel trace} \label{endomorphism_contributions}

We start computing all the terms in the expansion of \eqref{diagrams} involving solely the endomorphism $\U$ as external line.
We include in this set of diagrams also the diagram with no external lines at all, for simplicity.
We need all the vertices with up to two external $\U$ lines, that in components are
\begin{eqnarray}\label{endomorphism_vertices}
 \myincludegraphics{30}{0.45}{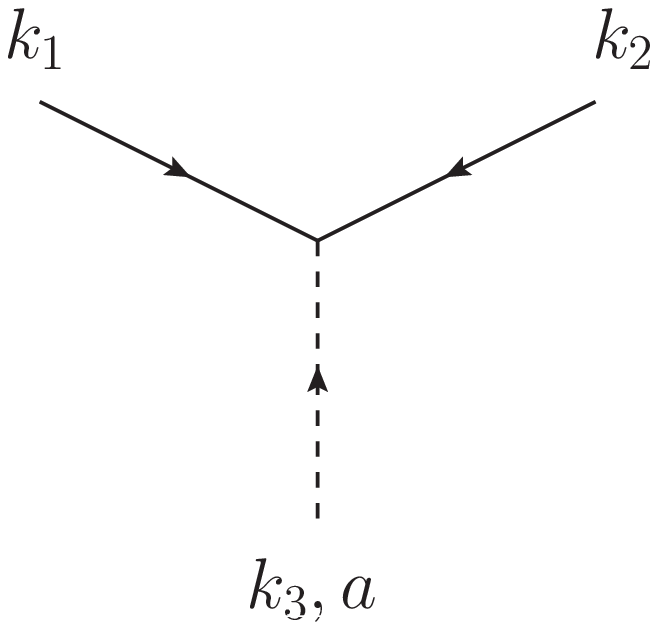}
 &=&\frac{\delta L}{\delta \phi_{k_1} \delta \phi_{k_2} \delta \U^a_{k_3}}[0,0]=T^a\,\\
 \myincludegraphics{30}{0.45}{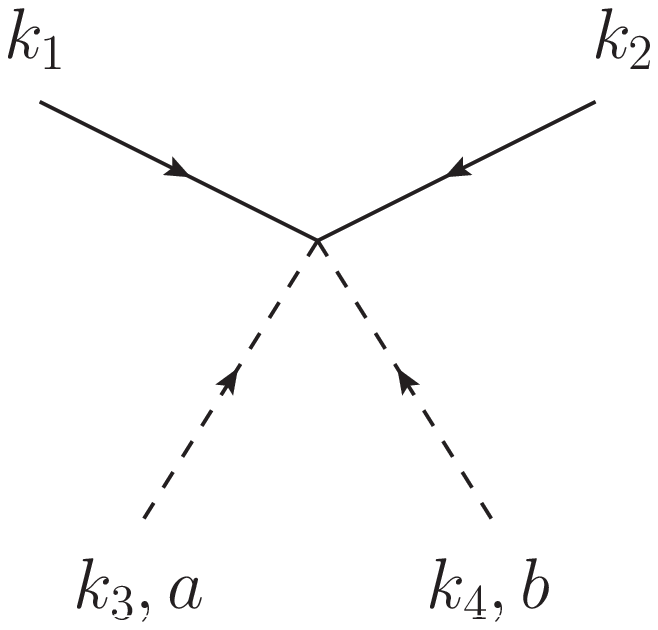}
 &=&\frac{\delta L}{\delta \phi_{k_1} \delta \phi_{k_2} \delta \U^a_{k_3}\delta\U^b_{k_4}}[0,0]=0\,.
\end{eqnarray}
We adopted dashed lines for the external endomorphism $\U$.

The first diagram we compute is the single loop with no external lines.
According to \eqref{diagrams} and using \eqref{general_propagator} in the simplest case of $n=0$,
in momentum space it is
\begin{eqnarray}
 \left.\Tr K^s\right|_{\Phi=0}
 &=&
 \myincludegraphics{14}{0.30}{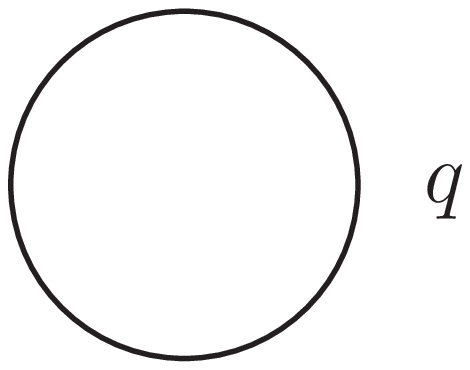}\nonumber\\
 &=&
 \int_q e^{-s q^2}\, \tr{\mathbf{1}}
 =
 \frac{1}{(2\pi)^d} \int d^dq \,e^{-s q^2}\, \tr{\mathbf{1}}\nonumber\\
 &=&
 \frac{\tr{\mathbf{1}}}{(4\pi s)^{d/2}}\,.
\end{eqnarray}
This has to be compared with the ansatz \eqref{ansatz1} in the collective limit $\Phi=0$ and therefore uniquely determines the constant $g_0=1$.

Then we compute the endomorphism $1$-point function. Due to momentum conservation, the incoming momentum of the endomorphism line must be zero. We obtain
\begin{eqnarray}\label{1pf_U}
 \left.\frac{\delta\Tr K^s}{\delta \U^a_0}\right|_{\Phi=0}
 &=&
 -s \myincludegraphics{14}{0.30}{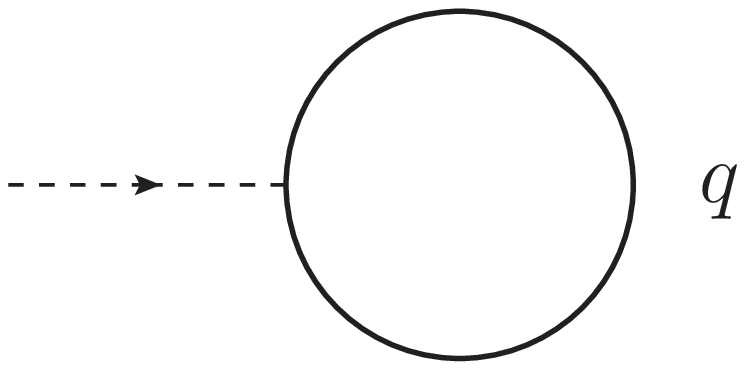}\nonumber\\
 &=&
 -\frac{s}{(2\pi)^d} \int d^dq \,e^{-s q^2}\,\tr T^a\nonumber\\
 &=&
 -s\frac{1}{(4\pi s)^{d/2}}\, \tr T^a\,.
\end{eqnarray}
Again, we can compare \eqref{1pf_U} with \eqref{ansatz1} and determine $g_{U,0}=-1$.

Finally, we want to obtain the endomorphism $2$-point function.
We observe that the vertex with two external $\U$ lines in \eqref{endomorphism_vertices} is zero,
therefore the expansion consists solely of one diagram
\begin{eqnarray}\label{2pf_UU}{}
 \left.\frac{\delta^2\Tr K^s}{\delta \U^a_p \delta \U^b_{-p}}\right|_{\Phi=0}
 &=&
 2 s^2 \myincludegraphics{22}{0.30}{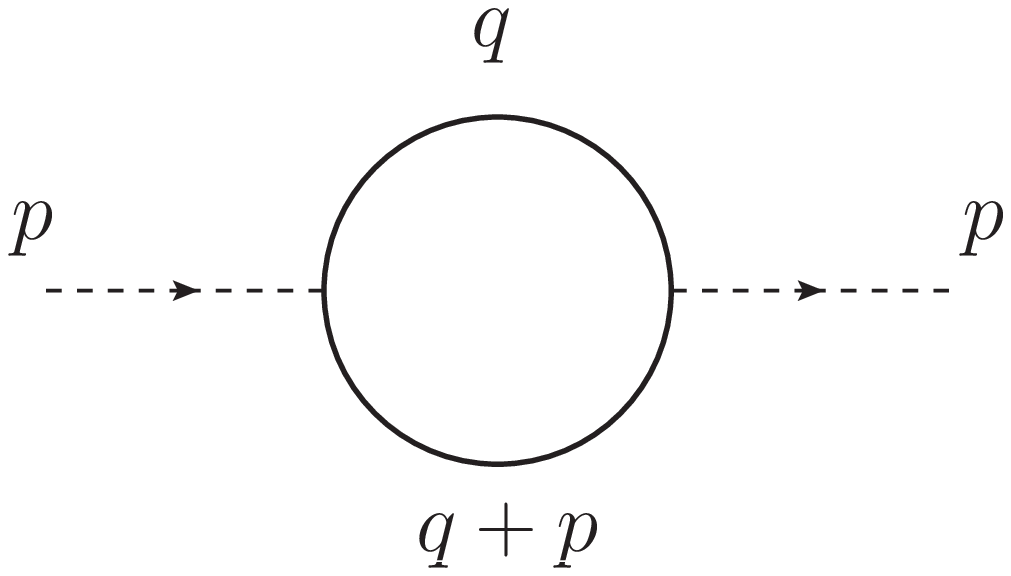}\nonumber
 \\
 {}
 &=&
 2 s^2 \int_0^1 dt_1 \int_0^{t_1} dt_2 \int_q\,e^{-s (1-t_1+t_2)q^2-s(t_1-t_2)(q+p)^2} \tr T^{(a}T^{b)}\, ,
\end{eqnarray}
where momentum conservation has been used.
It is convenient to introduce the new variable $\xi=t_1-t_2$ and rewrite the exponential of \eqref{2pf_UU} as
\begin{eqnarray}
 (1-t_1+t_2)q^2+(t_1-t_2)(q+p)^2
 &=&
 (q+\xi p)^2+\xi(1-\xi)p^2\,.
\end{eqnarray}
We can thus shift the momentum variable inside the integral of \eqref{2pf_UU}
and change the parametric integrals into a $\xi$-integration, that gives an additional factor of $1/2$.
These manipulations give
\begin{eqnarray}\label{2pf_UU2}
 \left.\frac{\delta^2\Tr K^s}{\delta \U^a_p \delta \U^b_{-p}}\right|_{\Phi=0}
 &=&
 \frac{s^2}{(4 \pi s)^{d/2}} f(s p^2) \tr T^{(a}T^{b)}\,,
\end{eqnarray}
where we defined the function
\begin{eqnarray}\label{basic_function}
 f(x)
 &\equiv&
 \int_0^1 d\xi e^{-\xi(1-\xi)x}
\end{eqnarray}
which is the basic form factor.
The relation \eqref{2pf_UU2} has to be compared to the second functional derivative
with respect to $\U$ of the ansatz \eqref{ansatz1}, that is
\begin{eqnarray}\label{2pf_UU3}
 \left.\frac{\delta^2\Tr K^s}{\delta \U^a_p \delta \U^b_{-p}}\right|_{\Phi=0}
 &=&
 \frac{2 s^2}{(4 \pi s)^{d/2}} f_U(s p^2) \tr T^{(a}T^{b)}\,.
\end{eqnarray}
Equating \eqref{2pf_UU2} and \eqref{2pf_UU3} we obtain
\begin{eqnarray}\label{fU}
 f_U(x) &=& \frac{1}{2} f(x)\,.
\end{eqnarray}

\subsection{Connection contribution to the heat kernel trace} \label{connection_contributions}

In this section we want to exploit all the terms of the expansion of \eqref{diagrams} involving external lines of the connection $A_\mu$.
We adopt a curly notation for these external lines. The vertices we need in this subsection are
\begin{eqnarray}\label{connection_vertices}
 \myincludegraphics{30}{0.45}{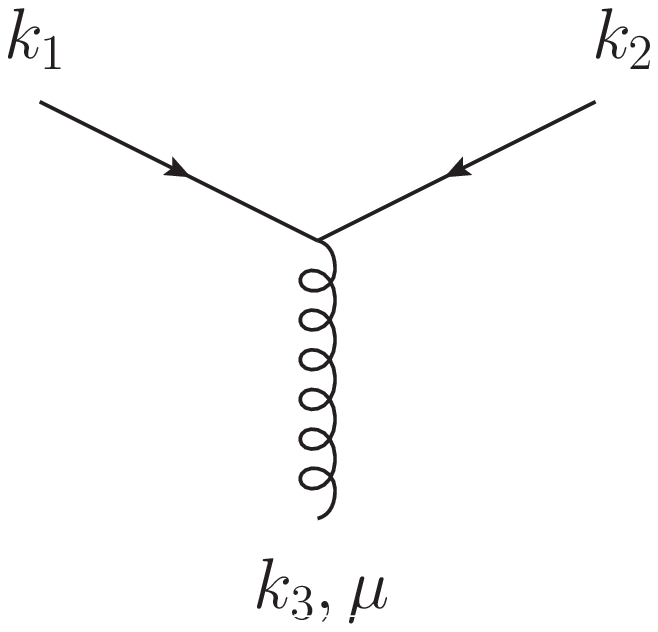}
 &=&\frac{\delta L}{\delta \phi_{k_1} \delta \phi_{k_2} \delta A^a_{k_3;\mu}}[0,0]=(k_1-k_2)^\mu T^a \,,\\
 \myincludegraphics{30}{0.45}{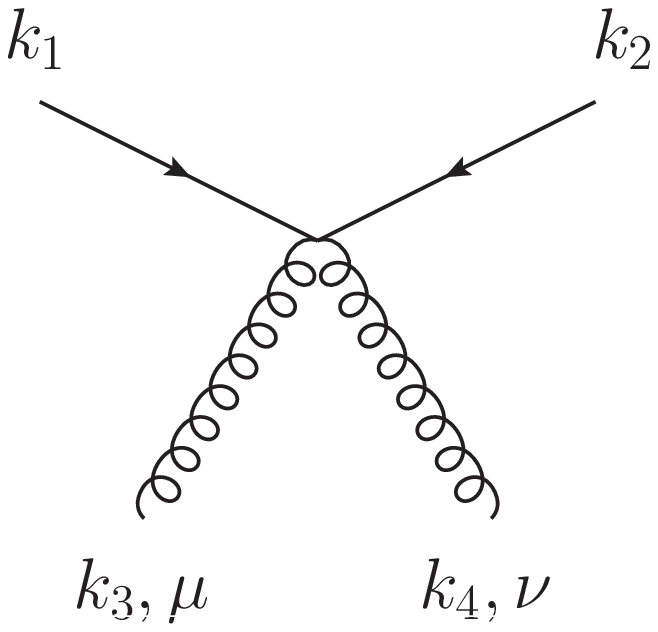}
 &=&\frac{\delta L}{\delta \phi_{k_1} \delta \phi_{k_2} \delta A^a_{k_3;\mu}\delta A^b_{k_4;\nu}}[0,0]=2 \delta^{\mu\nu} \,\tr T^{(a}T^{b)}\,,
\end{eqnarray}
where the parentheses imply symmetrization $T^{(a}T^{b)}=1/2(T^aT^b+T^bT^a)$.

It is possible to compute the $1$-point function of the expansion \eqref{diagrams} involving an external $A_\mu$ line
and show that it is zero. This agrees with the fact that no term linear in $A_\mu$ is present in the ansatz.
Since this computation is not particularly instructive, we shall not show it here.

We therefore turn our attention directly to the $2$-point function. It consists of two diagrams
\begin{eqnarray}\label{2pf_AA}
 \left.\frac{\delta^2\Tr K^s}{\delta A^a_{p;\mu}\delta A^b_{-p;\nu}}\right|_{\Phi=0}
 &=&
 2 s^2 \myincludegraphics{13}{0.30}{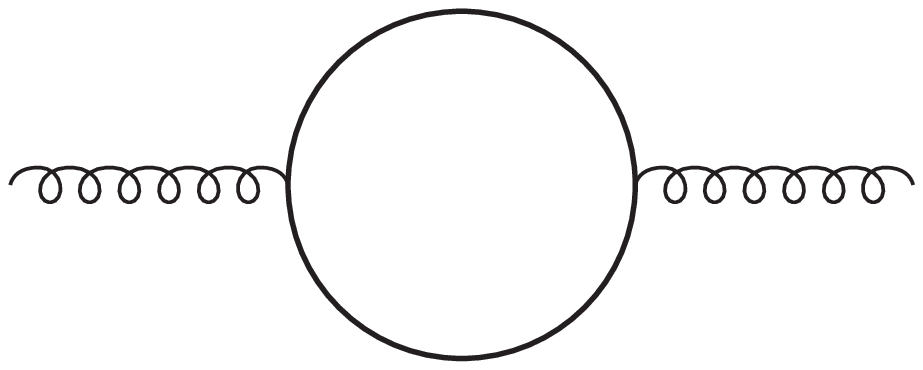} -s \myincludegraphics{17}{0.30}{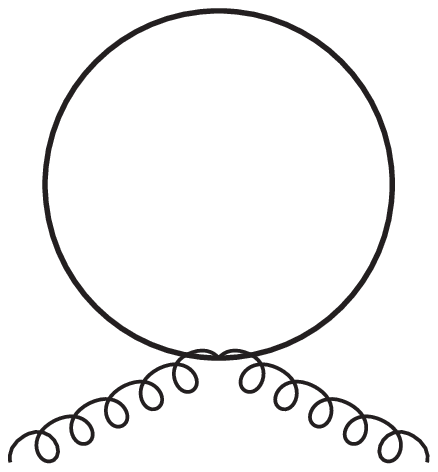}\,,
\end{eqnarray}
that we compute separately.
By the use of \eqref{connection_vertices}, the first diagram of \eqref{2pf_AA} gives
\begin{eqnarray}{}
 \myincludegraphics{13}{0.30}{images/2pf_sunset_AA.eps}
 &=&
 \int dt_1\, dt_2
 \int_q (2q+p)^\mu (2q+p)^\nu e^{-s (1-t_1+t_2)q^2-s(t_1-t_2)(q+p)^2} \tr T^{(a}T^{b)}\,.
 \nonumber
\end{eqnarray}
We perform the same manipulations that we did for the computation of \eqref{2pf_UU} and subsequently reduce the tensor structure inside the integration thanks to rotational invariance.
The result is
\begin{eqnarray}{}\label{2pf_AA_sunset}
 \myincludegraphics{13}{0.30}{images/2pf_sunset_AA.eps}
 &=&
 \frac{1}{(4\pi s)^{d/2}}\int_0^1d\xi\, e^{-s\xi(1-\xi)p^2}\left\{ \frac{\delta^{\mu\nu}}{s} + \frac{(1-2\xi)^2}{2} p^\mu p^\nu\right\}
 \tr T^{(a}T^{b)}
 \nonumber
 \\
 {}
 &=&
 \frac{1}{(4\pi s)^{d/2}}\left\{ \frac{f(sp^2)}{s}\delta^{\mu\nu} + \frac{1-f(sp^2)}{sp^2} p^\mu p^\nu\right\}
 \tr T^{(a}T^{b)}\,,
\end{eqnarray}
where in the second line we used the fact that the basic form factor \eqref{basic_function} has the property
\begin{eqnarray}
 \frac{1}{2}\int_0^1 d\xi\, e^{-\xi(1-\xi)x} (1-2\xi)^2
 &=&
 -\frac{1}{x}[f(x)-1]\,.
\end{eqnarray}
The second diagram of \eqref{2pf_AA} is much easier to compute and gives
\begin{eqnarray}\label{2pf_AA_tadpole}
 \myincludegraphics{17}{0.30}{images/2pf_tadpole_AA.eps}
 &=&
 2 \delta^{\mu\nu} \int_q \,e^{-s q^2}
 \tr T^{(a}T^{b)}
 \nonumber
 \\
 &=&
 \frac{2}{(4\pi s)^{d/2}} \delta^{\mu\nu}
 \tr T^{(a}T^{b)}\,.
\end{eqnarray}
Summing up \eqref{2pf_AA_sunset} and \eqref{2pf_AA_tadpole} with the correct overall factors, we determine \eqref{2pf_AA} as
\begin{eqnarray}\label{2pf_AA2}
 \left.\frac{\delta^2\Tr K^s}{\delta A^a_{p;\mu}\delta A^b_{-p;\nu}}\right|_{\Phi=0}
 &=&
 -2\frac{s^2}{(4\pi s)^{d/2}}p^2 P_T^{\mu\nu}
 \frac{1-f(s p^2)}{sp^2} \tr T^{(a}T^{b)}\,.
\end{eqnarray}
where we introduced the transverse projector $P_T^{\mu\nu} $.
This and the longitudinal projectors are functions of the incoming momentum $p_\mu$ and defined as
\begin{eqnarray}\label{vector_projectors}
 P_T^{\mu\nu} \equiv \delta^{\mu\nu} - \frac{p^\mu p^\nu}{p^2}\,,
 &\qquad\qquad&
 P_L^{\mu\nu} \equiv \frac{p^\mu p^\nu}{p^2}\,.
\end{eqnarray}

The hessian of the ansatz \eqref{ansatz1} is also transverse and is easily computed as
\begin{eqnarray}\label{2pf_AA3}
 \left.\frac{\delta^2\Tr K^s}{\delta A^a_{p;\mu}\delta A^b_{-p;\nu}}\right|_{\Phi=0}
 &=&
 -4\frac{s^2}{(4\pi s)^{d/2}}p^2 P_T^{\mu\nu} f_\Omega(sp^2) \tr T^{(a}T^{b)}\,.
\end{eqnarray}
A trivial comparison of \eqref{2pf_AA2} and \eqref{2pf_AA3}
gives
\begin{eqnarray}\label{fOmega}
 f_\Omega(x) &=& -\frac{1}{2x}[f(x)-1]\,.
\end{eqnarray}

\subsection{Metric contribution to the heat kernel trace}\label{gravity_contributions}

In this subsection we complete the computation of the form factors appearing in the ansatz \eqref{ansatz1},
including all the diagrams of \eqref{diagrams} involving at least one external graviton line.
We need to compute the vertices of \eqref{HK_PT_11} involving solely graviton external lines
\begin{eqnarray}\label{graviton_vertices}
 \myincludegraphics{30}{0.45}{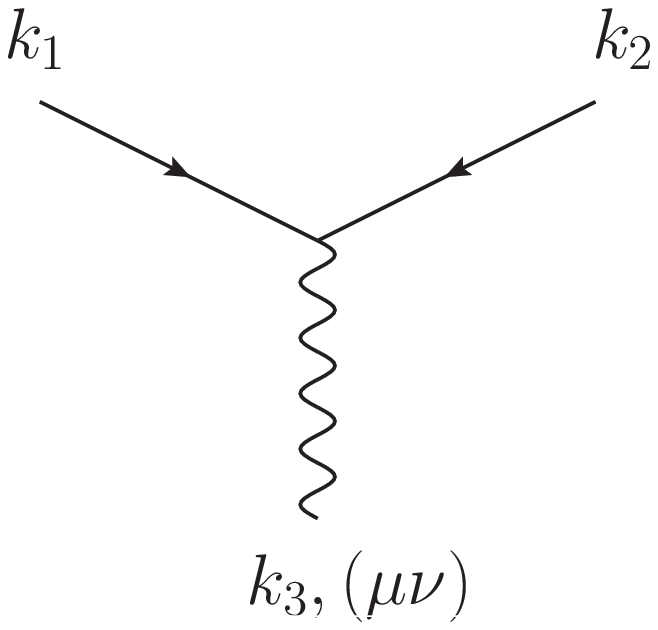}
 &=&
 k_1^{(\mu}k_2^{\nu)}-\frac{1}{4}(k_1+k_2)^2 \delta^{\mu\nu}
 \\
 \myincludegraphics{30}{0.45}{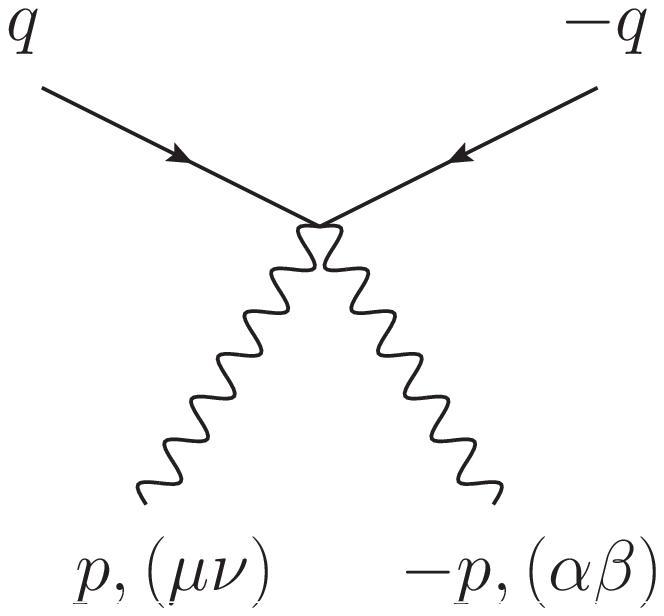}
 &=&
 2q^{(\mu} g^{\nu)(\alpha} q^{\beta)}
 +\frac{1}{4} p^2 \delta^{\mu\nu} \delta^{\alpha\beta}\,,
\end{eqnarray}
where in the second one we do not give the most general dependence on the incoming momenta for simplicity.
The first vertex of \eqref{graviton_vertices} is derived from \eqref{HK_PT_11}, by considering all the terms linear in $h_{\mu\nu}$ when $g_{\mu\nu}=\delta_{\mu\nu}+h_{\mu\nu}$;
similarly, the second vertex comes from all the terms quadratic in $h_{\mu\nu}$.
In this section we will also need the endomorphism vertices given in \eqref{endomorphism_vertices}.
In principle, we would also need vertices mixing the endomorphism and graviton lines, however,
due to the density normalization of the basis \eqref{basis2} we adopt, the endomorphism is decoupled from the metric in the Laplacian action \eqref{HK_PT_11}
and therefore no mixed vertex is present.
Further, since the ansatz \eqref{ansatz1} does not possess any form-factor involving the curvature $\Omega_{\mu\nu}$
and any of the tensors $R$, $R_{\mu\nu}$ or $R_{\mu\nu\alpha\beta}$, there is no need to report the vertex involving
one external metric and one external connection lines. It is easy to see that, in this case, the $2$-point function
of \eqref{diagrams} is identically zero.
In fact, the computation of the $2$-point function of the heat kernel with external $A_\mu$ and $h_{\mu\nu}$ lines can be used as a check of the consistency of the ansatz.
In the case of the trace of the heat kernel the result is zero, that implies that no form factor relates $\Omega_{\mu\nu}$ with $R$ or $R_{\mu\nu}$.
Since this computation would not determine any of the form factors of the ansatz \eqref{ansatz1}, we shall not include it.

Let us proceed computing the $1$-point graviton function of \eqref{diagrams}
\begin{eqnarray}\label{1pf_h}
 \left.\frac{\delta\Tr K^s}{\delta h_{0;\mu\nu}}\right|_{\Phi=0}
 &=&
 -s \myincludegraphics{14}{0.30}{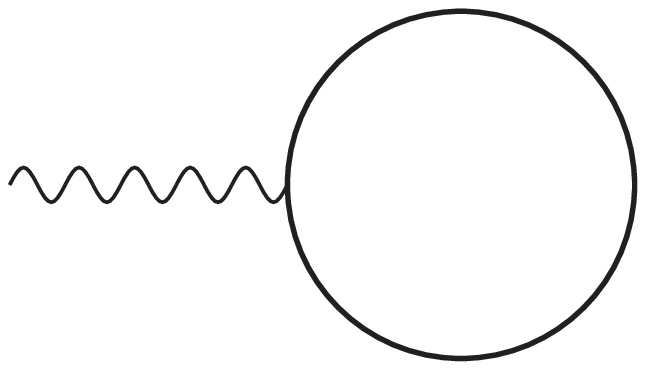}\,
 \nonumber\\
 &=& \frac{\tr\mathbf{1}}{2(4\pi s)^{d/2}}\,.
\end{eqnarray}
This has to be compared with the functional derivative of the ansatz \eqref{ansatz1} in the same limit.
It is easy to see that the result is consistent with the determination $g_0=1$.
However \eqref{1pf_h} cannot be used to compute $g_{R,0}$. This can be seen both from the ansatz and from \eqref{1pf_h}.
On the ansatz side, one shows easily that the first order expansion in the metric fluctuation $h_{\mu\nu}$
around a flat background of the term of \eqref{ansatz1} corresponding to $g_{R,0}$ is a total derivative and therefore vanishes
upon integration for our boundary conditions. On the side of \eqref{1pf_h}, we observe that momentum conservation forces
the incoming momentum of the $h_{\mu\nu}$ external line to be zero and $g_{R,0}$ cannot be detected since it would appear with a quadratic power of the incoming momentum.
The correct way to compute $g_{R,0}$ in our formalism is by considering an ansatz for the heat kernel, rather one for the trace of the heat kernel.
This will be shown in \ref{appendixB}.
For the moment, let us assume the well known value from the literature $g_{R,0}=1/6$, that we are going to need in the following.

We now move to the computation of the form factors of \eqref{ansatz1} involving at least one curvature tensor constructed with the metric.
The $2$-point graviton function is
\begin{eqnarray}\label{2pf_hh}
 \left.\frac{\delta^2\Tr K^s}{\delta h_{p;\mu\nu}\delta h_{-p;\alpha\beta}}\right|_{\Phi=0}
 &=&
 2 s^2 \myincludegraphics{13}{0.30}{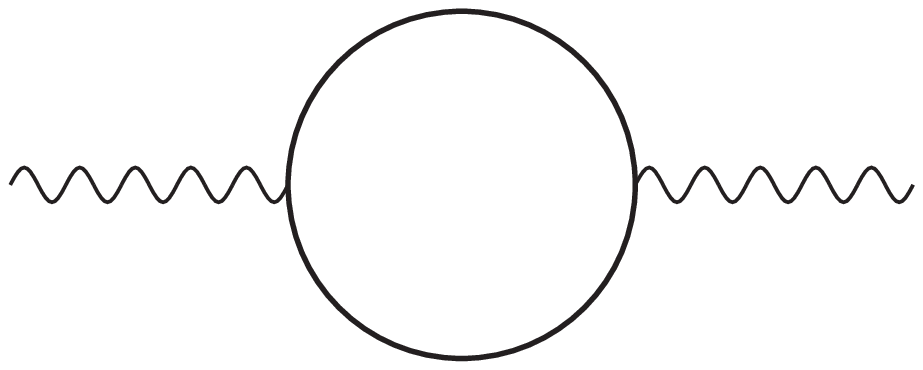} -s \myincludegraphics{17}{0.30}{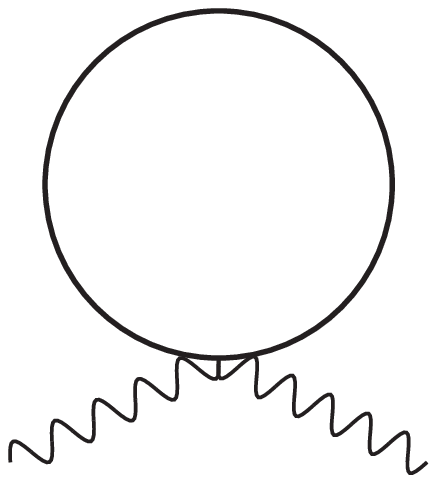}\,.
\end{eqnarray}
The
evaluation of the diagrams goes as in the previous cases, we find
the following result:
\begin{eqnarray}\label{2pf_hh2}{}
 \left.\frac{\delta^{2}\Tr  K^{s}}{\delta h_{p;\mu\nu}\delta h_{-p;\alpha\beta}}\right|_{\Phi=0}
 & = &
 \frac{1}{(4\pi s)^{d/2}}\Bigl\{ \Bigl[-1+\frac{1}{2}f(sp^{2})\Bigr] P_{2}^{\mu\nu,\alpha\beta}+\Bigl[-1-\frac{d-1}{8}sp^{2}\nonumber \\
 {}
 &  &
 +\frac{1}{16}(4(d+1)+4(d-1)sp^{2}+(d-1)s^{2}p^{4})f(sp^{2})\Bigr]P_{S}^{\mu\nu,\alpha\beta}\nonumber \\
 {}
 &  &
 -\frac{1}{2}P_{1}^{\mu\nu,\alpha\beta}-\frac{1}{4}P_{\sigma}^{\mu\nu,\alpha\beta}
 +\frac{\sqrt{d-1}}{4}\Bigl(P_{S\sigma}^{\mu\nu,\alpha\beta}+P_{\sigma S}^{\mu\nu,\alpha\beta}\Bigr)\Bigr\} \,.
\end{eqnarray}
In \eqref{2pf_hh2} we have introduced the following set of six generalized tensor projectors \cite{percacci}:
\begin{eqnarray}\label{tensor_projectors}
 P_{2;\mu\nu,\alpha\beta}
 & = &
 \frac{1}{2}\left(P^T_{\mu\alpha}P^T_{\nu\beta} + P^T_{\mu\beta}P^T_{\nu\alpha}\right)
 -\frac{1}{d-1}P^T_{\mu\nu}P^T_{\alpha\beta}
 \nonumber \\
 P_{1;\mu\nu,\alpha\beta}
 & = &
 \frac{1}{2}\left[P^T_{\mu\alpha}P^L_{\nu\beta}+P^T_{\mu\beta}P^L_{\nu\alpha}
 +P^T_{\nu\alpha}P^L_{\mu\beta}+P^T_{\nu\beta}P^L_{\mu\alpha}\right]
 \nonumber \\
 P_{S;\mu\nu,\alpha\beta}
 & = &
 \frac{1}{d-1}P^T_{\mu\nu}P^T_{\alpha\beta}
 \nonumber \\
 P_{S\sigma;\mu\nu,\alpha\beta}
 & = &
 \frac{1}{\sqrt{d-1}}P^T_{\mu\nu}P^L_{\alpha\beta}
 \nonumber \\
 P_{\sigma S;\mu\nu,\alpha\beta}
 & = &
 \frac{1}{\sqrt{d-1}}P^L_{\mu\nu}P^T_{\alpha\beta}
 \nonumber \\
 P_{\sigma;\mu\nu,\alpha\beta}
 & = &
 P^L_{\mu\nu}P^L_{\alpha\beta}\,.
\end{eqnarray}
We list here some of their properties:
\begin{eqnarray}\label{projector_properties}
 \mathbf{P}_{i}^{2} &=& \mathbf{P}_{i}\,, \qquad{\rm for}\,\, i=2,1,S,\sigma
 \nonumber\\
 (\mathbf{P}_{S\sigma}+\mathbf{P}_{\sigma S})^2 &=& \mathbf{P}_{S}+\mathbf{P}_{\sigma}
 \,,
\end{eqnarray}
that are useful when computing the components of \eqref{2pf_hh2}.
The Hessian of \eqref{ansatz1} is also computed
\begin{eqnarray}\label{2pf_hh3}{}
 \left.\frac{\delta^{2}\Tr  K^{s}}{\delta h_{p;\mu\nu}\delta h_{-p;\alpha\beta}}\right|_{\Phi=0}
 & = &
 \frac{1}{(4\pi s)^{d/2}}\Bigl\{ \Bigl[-\frac{g_{0}}{2}-\frac{g_{R,0}}{2}sp^{2}+\frac{1}{2}s^{2}p^{4}f_{Ric}\left(sp^{2}\right)\Bigr]P_{2}^{\mu\nu,\alpha\beta}
 \nonumber \\
 {}
 &  & +\Bigl[\frac{(d-3)g_{0}}{4}+\frac{(d-2)g_{R,0}}{2}sp^{2}
 \nonumber \\
 {}
 &  & +\frac{d}{2}s^{2}p^{4}f_{Ric}(sp^{2})+2(d-1)s^{2}p^{4}f_{R}(sp^{2})\Bigr]P_{S}^{\mu\nu,\alpha\beta}
 \nonumber \\
 {}
 &  & -\frac{g_{0}}{2}P_{1}^{\mu\nu,\alpha\beta}-\frac{g_{0}}{4}P_{\sigma}^{\mu\nu,\alpha\beta}+\frac{\sqrt{d-1}g_{0}}{4}
 \Bigl(P_{S\sigma}^{\mu\nu,\alpha\beta}+P_{\sigma S}^{\mu\nu,\alpha\beta}\Bigr)\Bigr\}
 \,.
\end{eqnarray}
Now we use the fact that $g_0=1$ as previously computed, and that $g_{R,0}=1/6$ as will be shown in appendix \ref{appendixB}.
Equating \eqref{2pf_hh2} and \eqref{2pf_hh3}, we can solve the system of the coefficients of the projectors in terms of the form factors
$f_R(x)$ and $f_{Ric}(x)$.
We find that
\begin{eqnarray}
 \label{fRic+R}
 f_{Ric}(x)
 & = &
 \frac{1}{6x}+\frac{1}{x^{2}}\bigl[f(x)-1\bigr]
 \nonumber \\
 f_{R}(x)
 & = &
 \frac{1}{32}f(x)+\frac{1}{8x}f(x)-\frac{7}{48x}-\frac{1}{8x^{2}}\bigl[f(x)-1\bigr]\,.
\end{eqnarray}

We finally want to compute the last form factor of the ansatz \eqref{ansatz1}, namely $f_{UR}(x)$.
To this end we first calculate the $2$-point function of the heat kernel with external $\U$ and $h_{\mu\nu}$ lines,
that is simplified by the fact that the endomorphism does not couple to the metric explicit in the Laplacian action \eqref{HK_PT_11}.
There is only one diagram contributing,
\begin{eqnarray}\label{2pf_hU1}
 \left.\frac{\delta^{2}\Tr  K^{s}}{\delta h_{p;\mu\nu}\delta \U^a_{-p}}\right|_{\Phi=0}
 &=&
 2 s^2 \myincludegraphics{13}{0.30}{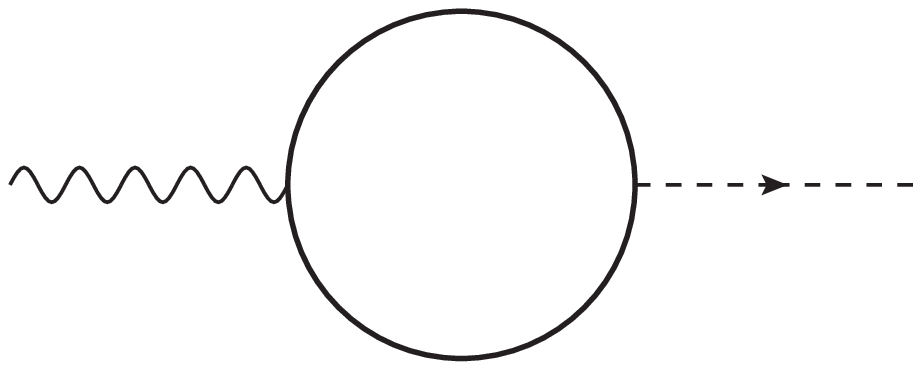}\nonumber
 \\
 &=&
 \frac{1}{(4\pi s)^{d/2}}\Bigl[-\frac{s}{2}P_L^{\mu\nu}-\frac{1}{4}(2s+s^{2}p^{2})P_T^{\mu\nu}f(sp^{2})\Bigr] \tr T^{a}\,,
\end{eqnarray}
that has to be compared with the corresponding Hessian of the ansatz \eqref{ansatz1}
\begin{eqnarray}\label{2pf_hU2}{}
 \left.\frac{\delta^{2}\Tr  K^{s}}{\delta h_{p;\mu\nu}\delta \U^a_{-p}}\right|_{\Phi=0}
 &=&
 \frac{1}{(4\pi s)^{d/2}}\Bigl[\frac{g_{U,0}s}{2}P_L^{\mu\nu}+\Bigl(\frac{g_{U,0}s}{2}+s^{2}p^{2}f_{RU}(sp^{2})\Bigr)P_T^{\mu\nu}\Bigr] \tr T^{a}\,.
\end{eqnarray}
Equating \eqref{2pf_hU1} and \eqref{2pf_hU2} one easily confirms that $g_{U,0}=-1$ and that
\begin{eqnarray} 
 \label{fUR}
 f_{RU}(x)
 & = &
 -\frac{1}{4}f(x)-\frac{1}{2x}[f(x)-1]\,.
\end{eqnarray}

\section{Results for the non-local expansion}

In this section we briefly collect all the results for the non-local heat kernel expansion,
analyze their properties
and compare them with the existing literature.

\begin{figure}[t]
 \centering{}
 \includegraphics[scale=0.85]{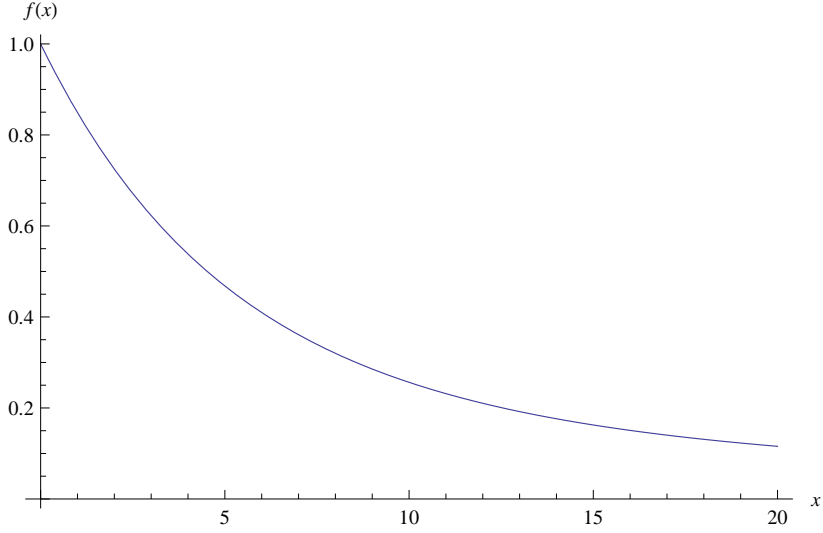}
 \caption{\small{Plot of the basic form factor $f(x)$. The large $x$ behavior shows an exponential tail, that cannot be obtained with finitely many terms in a small-$x$ (Seeley-deWitt) expansion.}}
 \label{plot_f}
\end{figure}

The expansion of the trace of the heat-kernel reads as follows
\begin{eqnarray} \label{HK_2.2}{}
 \Tr  K^{s}
 & = &
 \frac{1}{(4\pi s)^{d/2}}
 \int d^{d}x\sqrt{g}\,\tr\Bigl\{
 \mathbf{1}
 -s\mathbf{U}
 +s\mathbf{1}\frac{R}{6}
 +s^{2}\Bigl[
  \mathbf{1}R_{\mu\nu}f_{Ric}(s\Box)R^{\mu\nu}
  +\mathbf{1}R\, f_{R}(s\Box)R
  \nonumber \\
 {}
 &  &
  +R\, f_{RU}(s\Box)\U
  +\U f_{U}(s\Box)\U
  +\Omega_{\mu\nu}f_{\Omega}(s\Box)\Omega^{\mu\nu}\Bigr]
  +O(\mathcal{R}^{3})\Bigr\} \,,
\end{eqnarray}
where $\mathcal{R}$ represents the endomorphism or any of the curvatures and $\Box=-D^2$.
For the heat kernel form factors in \eqref{HK_2.2}, we collect the results of \eqref{fU}, \eqref{fOmega}, \eqref{fRic+R} and \eqref{fUR}.
These have been found to be
\begin{eqnarray}\label{HK_2.21}
 f_{Ric}(x) & = & \frac{1}{6x}+\frac{1}{x^{2}}\left[f(x)-1\right]\nonumber \\
 f_{R}(x) & = & \frac{1}{32}f(x)+\frac{1}{8x}f(x)-\frac{7}{48x}-\frac{1}{8x^{2}}\left[f(x)-1\right]\nonumber \\
 f_{RU}(x) & = & -\frac{1}{4}f(x)-\frac{1}{2x}\left[f(x)-1\right]\nonumber \\
 f_{U}(x) & = & \frac{1}{2}f(x)\nonumber \\
 f_{\Omega}(x) & = & -\frac{1}{2x}\left[f(x)-1\right]\,,
\end{eqnarray}
and all depend on the basic heat kernel form factor $f(x)$ that is defined
in terms of a parameter integral
\begin{equation}\label{HK_2.3}
 f(x)=\int_{0}^{1}d\xi\, e^{-x\xi(1-\xi)}\,.
\end{equation}
Our findings are in complete agreement with \cite{Barvinsky_Vilkovisky_1987_1990}, but given in a different curvature basis as we will show in subsection \ref{BVbasis}.
We observe that \eqref{HK_2.2} is universal, in the sense that it depends on the dimensionality
$d$ of the manifold only through the overall factor $(4\pi s)^{-d/2}$. This is not a feature that can be expected
from the heat kernel of any operator, but rather is a property of considering a Laplace-like operator.
A plot of the basic form factor \eqref{HK_2.3} is given in figure \ref{plot_f}.

Using in \eqref{HK_2.21} the Taylor expansion of the basic form factor \eqref{HK_2.3}
\begin{equation}f(x)=1-\frac{x}{6}+\frac{x^{2}}{60}+O(x^{3})\,,\end{equation}
gives the following short time expansion for the form factors
\begin{eqnarray}\label{HK_2.22}
 f_{Ric}(x) & = & \frac{1}{60}-\frac{x}{840}+\frac{x^{2}}{15120}+O(x^{3})\nonumber \\
 f_{R}(x) & = & \frac{1}{120}-\frac{x}{336}+\frac{11x^{2}}{30240}+O(x^{3})\nonumber \\
 f_{RU}(x) & = & -\frac{1}{6}+\frac{x}{30}-\frac{x^{2}}{280}+O(x^{3})\nonumber \\
 f_{U}(x) & = & \frac{1}{2}-\frac{x}{12}+\frac{x^{2}}{120}+O(x^{3})\nonumber \\
 f_{\Omega}(x) & = & \frac{1}{12}-\frac{x}{120}+\frac{x^{2}}{1680}+O(x^{3})\,.
\end{eqnarray}
Inserting \eqref{HK_2.22} into \eqref{HK_2.2} it is possible to recover the results of
the local expansion of the heat-kernel, for example those obtained with the Seeley-deWitt method.

However, having access to the full form factors \eqref{HK_2.21},
we can study the late time behavior of the heat kernel as well.
This is of interest, since by definition it is not accessible from the local expansion,
unless the resummation of infinitely many terms in performed.
From the late time expansion of the basic form factor
\begin{equation}
 f(x)=\frac{2}{x}+\frac{4}{x^{2}}+O\left(\frac{1}{x^{3}}\right)\,,
\end{equation}
we find
\begin{eqnarray} \label{HK_2.26}
 f_{Ric}(x) & = & \frac{1}{6x}-\frac{1}{x^{2}}+O\left(\frac{1}{x^{3}}\right)\nonumber \\
 f_{R}(x) & = & -\frac{1}{12x}+\frac{1}{2x^{2}}+O\left(\frac{1}{x^{3}}\right)\nonumber \\
 f_{RU}(x) & = & -\frac{2}{x^{2}}-\frac{8}{x^{3}}+O\left(\frac{1}{x^{4}}\right)\nonumber \\
 f_{U}(x) & = & \frac{1}{x}+\frac{2}{x^{2}}+O\left(\frac{1}{x^{3}}\right)\nonumber \\
 f_{\Omega}(x) & = & \frac{1}{2x}-\frac{1}{2x^{2}}+O\left(\frac{1}{x^{3}}\right)\,.
\end{eqnarray}

For some application of the form factors \eqref{HK_2.21}
and the corresponding expansions \eqref{HK_2.22} and \eqref{HK_2.26},
we refer to \cite{Barvinsky_Vilkovisky_1987_1990,Avramidi_1990_2002}.
Some comment on very important applications of \eqref{HK_2.21} in quantum field theory computations
will be given in subsection \ref{Weyl_basis}, when a different curvature basis will be adopted.

\subsection{The two dimensional limit}

The two dimensional limit of \eqref{HK_2.2} is rather interesting and deserves a separate study.
For $d=2$ the relation $R_{\mu\nu}=\frac{1}{2}g_{\mu\nu}R$ holds and therefore the form factors
$f_R(x)$ and $f_{Ric}(x)$ combine. Neglecting endomorphism $\U$ and connection, we have that
\begin{eqnarray}\label{HK_2.23}
 \Tr K^{s}\bigl|_{d=2}
 &=&
 \frac{1}{4\pi s}\int d^{2}x\sqrt{g}\,\tr\mathbf{1}\Bigl[1+s\frac{R}{6}+s^{2}R\, f_{R2d}(s\Box)R+O(R^{3})\Bigr]\,,
\end{eqnarray}
where there is now only one form factor $f_{R2d}(x)$ defined as
\begin{eqnarray}\label{HK_2.24}
 f_{R2d}(x)
 &=&
 \frac{1}{32}f(x)+\frac{1}{16x}\bigl[2f(x)-1\bigr]+\frac{3}{8x^{2}}\bigl[f(x)-1\bigr]\,.
\end{eqnarray}
The late time expansion of \eqref{HK_2.24} is found to be
\begin{equation}\label{HK_2.27}
 f_{R2d}(x)=\frac{2}{x^{3}}+\frac{12}{x^{4}}+O\left(\frac{1}{x^{5}}\right)\,;
\end{equation}
we observe that in \eqref{HK_2.27} both the coefficients of $x^{-1}$ and $x^{-2}$ are zero \cite{Barvinsky_Vilkovisky_1987_1990}. 

\subsection{The Weyl basis}\label{Weyl_basis}

It is possible to change the basis $\{R_{\mu\nu},R,\U,\Omega_{\mu\nu}\}$ of the curvature invariants in \eqref{HK_2.2} to the set $\{C_{\mu\nu\alpha\beta},R,\U,\Omega_{\mu\nu}\}$,
where $C_{\mu\nu\alpha\beta}$ is the Weyl tensor \cite{Barvinsky:1995it}.
The expansion \eqref{HK_2.2} becomes
\begin{eqnarray}\label{HK_2.2201}{}
\Tr K^{s}
 & = &
 \frac{1}{(4\pi s)^{d/2}}\int d^{d}x\sqrt{g}\,\tr
 \Bigl\{ \mathbf{1}-s\U+s\mathbf{1}\frac{R}{6}+s^{2}\Bigl[\mathbf{1}C_{\mu\nu\alpha\beta}f_{C}(s\Box)C^{\mu\nu\alpha\beta}\nonumber \\
 {}
 &  &
 +\mathbf{1}R\, f_{Rbis}(s\Box)R+R\, f_{RU}(s\Box)\U+\mathbf{U}f_{U}(s\Box)\U\nonumber \\
 {}
 &  &
 +\Omega_{\mu\nu}f_{\Omega}(s\Box)\Omega^{\mu\nu}\Bigr]+O(\mathcal{R}^{3})\Bigr\} \,.
\end{eqnarray}
We defined two new form factors
\begin{eqnarray}\label{HK_2.221}
 f_{C}(x) & = & \frac{d-2}{4(d-3)}f_{Ric}(x)\nonumber \\
 f_{Rbis}(x) & = & \frac{d}{4(d-1)}f_{Ric}(x)+f_{R}(x)\,.\label{HK_2.2202}
\end{eqnarray}
The new basis spoils the universality property, because it induces an explicit
dependence on $d$ into the form factors. We refer to \ref{appendixA} for a more detailed derivation of \eqref{HK_2.221}.
The short time expansions of the form factors \eqref{HK_2.2201} are
$f_{C}(x)=(d-2)/360(d-3)+O(x)$ and $f_{Rbis}(x)=(3d-2)/240(d-1)+O(x)$.

The Weyl basis presented here is particularly useful in quantum field theory applications.
In fact, the Hessian
of the square of the Weyl tensor, when evaluated on flat space,
is proportional to the projector ${\mathbf P}_2$ and is then related to the inverse momentum-space propagator
of the irreducible spin-$2$ component of the fluctuation $h_{\mu\nu}$.
On a general background, the form factor $f_C(x)$ is thus directly related to the full quantum propagator of ``graviton'' excitations \cite{Codello}.

\subsection{The BV basis}\label{BVbasis}

We finally present the result \eqref{HK_2.2} in the basis of curvatures in which it was originally
obtained by Barvinsky and Vilkovisky \cite{Barvinsky_Vilkovisky_1987_1990}.
We therefore introduce a new basis, that we refer to as the BV basis, where the trace of the heat kernel
is expanded in powers of the set of generalized curvatures $\{R_{\mu\nu},R,\mathbf{P},\Omega_{\mu\nu}\}$ with $\mathbf{P}=-\U+\mathbf{1}R/6$.
In the new basis \eqref{HK_2.2} becomes
\begin{eqnarray} \label{HK_2.2bis}{}
 \Tr  K^{s}
 & = &
 \frac{1}{(4\pi s)^{d/2}}
 \int d^{d}x\sqrt{g}\,\tr\Bigl\{
 \mathbf{1}
 +s\mathbf{P}
 +s^{2}\Bigl[
  \mathbf{1}R_{\mu\nu}f_{1}(s\Box)R^{\mu\nu}
  +\mathbf{1}R\, f_{2}(s\Box)R
  \nonumber \\
 {}
 &  &
  +R\, f_{3}(s\Box)\mathbf{P}
  +\mathbf{P} f_{4}(s\Box) \mathbf{P}
  +\Omega_{\mu\nu}f_{5}(s\Box)\Omega^{\mu\nu}\Bigr]
  +O(\mathcal{R}^{3})\Bigr\} \,,
\end{eqnarray}
and a new set of form factors $f_i(x)$ for $i=1,\dots,5$ has been introduced.
These new functions are related to those of \eqref{HK_2.21} by a simple linear transformation
\begin{eqnarray}\label{BV_basis}
 f_{1}(x) & = & f_{Ric}(x)\nonumber\\
 f_{2}(x) & = & f_{R}(x)+\frac{1}{36}f_U(x)+\frac{1}{6}f_{UR}(x)\nonumber\\
 f_{3}(x) & = & -\frac{1}{3}f_{U}(x)-f_{UR}(x)\nonumber\\
 f_{4}(x) & = & f_{U}(x)\nonumber\\
 f_{5}(x) & = & f_{\Omega}(x)\,.
\end{eqnarray}
We refer to appendix \ref{appendixA} for a complete derivation of \eqref{BV_basis}.
Substituting \eqref{BV_basis} into \eqref{HK_2.21}, it is possible to check by direct comparison
that our results do coincide with those of \cite{Barvinsky_Vilkovisky_1987_1990}.

\section{Resolvent method}

In this section we introduce a slightly different formalism to handle
the heat kernel trace which is based on the resolvent method.
The method is meant to be a momentum space extension of \cite{Gusynin_1997},
capable of capturing the non-local features of the form factors
computed in the previous sections.
It is based on the following relation
\begin{eqnarray}\label{RM_1}
 e^{-sx}
 & = &
 \int_{\cal C}\frac{id\theta}{2\pi}e^{-s\theta}\frac{1}{x-\theta}\,,
\end{eqnarray}
where $\theta$ is a complex variable and ${\cal C}$ is a countour in the
complex plane having in its interior all the poles of the inverse
function $(x-\theta)^{-1}$ of the right hand side. From now on, all the formulas
have to be suitably analytically continued in $\theta$
as much as possible.
The relation \eqref{RM_1} is very useful since it allows to express
the exponential of a differential operator, i.e.
the heat kernel, in terms of the inverse of the differential operator $\Delta-\theta$
\begin{equation}\label{RM_2}
 \Tr K^{s}=\int_{\cal C}\frac{id\theta}{2\pi}e^{-s\theta}\Tr \frac{1}{\Delta-\theta}=\int_\theta\Tr \frac{1}{L^{(2;0)}-\theta}\,,
\end{equation}
where we abbreviated $\int_\theta \equiv \int_{\cal C} i d\theta e^{-s \theta}/2\pi$
and used the fact that $\Delta$ acts on tensors of weight $1/2$ inside the trace
to write $\Delta=L^{(2;0)}[\phi;\Phi]$.
The inverse $(\Delta-\theta)^{-1}$ is usually called resolvent of $\Delta$.
It is important that in \eqref{RM_2} the contour ${\cal C}$
contains in its interior all the poles of the resolvent.

The application of functional derivatives with respect to $\Phi$ on
the right hand side of \eqref{RM_2} is obtained by virtue of the relation
$\delta M^{-1}=-M^{-1}\delta M M^{-1}$, that is valid for any operator $M$.
In this framework, we observe that the appearance
of the time ordered product in the perturbative expansion of the heat
kernel given in equation \eqref{HK_PT_7} is due to the fact that
the operator $\Delta$ and its variation $\delta\Delta$ do not commute
in general, i.e. $[\Delta,\delta\Delta]\neq0$.
At the price of introducing the auxiliary complex integration over $\theta$
we gain an easier way to obtain equations for the $n$-point functions of the heat kernel trace.
One simply has
\begin{eqnarray}\label{RM_4}{}
 \frac{\delta\Tr  K^{s}}{\delta\Phi(x)}
 &=&
 -\int_\theta \int_{yzw} \tr \left(\frac{1}{L^{(2;0)}-\theta}\right)_{yz}L_{zwx}^{(2;1)}\left(\frac{1}{L^{(2;0)}-\theta}\right)_{wy}\label{RM_3}
 \nonumber\\
 {}
 \frac{\delta^{2}\Tr  K^{s}}{\delta\Phi(x')\delta\Phi(x)}
 & = &
 2\int_\theta \int_{yzwuv}
 \tr
 \left(\frac{1}{L^{(2;0)}-\theta}\right)_{yz}
 L_{zwx}^{(2;1)}
 \left(\frac{1}{L^{(2;0)}-\theta}\right)_{wu}
 L_{uvx'}^{(2;1)}
 \left(\frac{1}{L^{(2;0)}-\theta}\right)_{vy}
 \nonumber \\
 {}
 &  &
 -\int_\theta \int_{yzw}
 \tr
 \left(\frac{1}{L^{(2;0)}-\theta}\right)_{yz}
 L_{zwxx'}^{(2;2)}
 \left(\frac{1}{L^{(2;0)}-\theta}\right)_{wy}\,.
\end{eqnarray}
One can now set $\Phi=0$ in \eqref{RM_4} and move to momentum space
in order to obtain
a diagrammatics for the $n$-point functions of the heat
kernel trace of section \ref{section2} that is slightly different from
\eqref{diagrams}.
This diagrammatic shares the vertices of the one developed in the previous sections.
The difference, instead, lies in the fact that
propagators are closer to the standard
ones of quantum field theory and, for this application, have complex Euclidean
mass square $-\theta$.

For the sake of the application, we re-obtain the results of subsection \ref{connection_contributions}
in this new formalism. For simplicity, we restrict our attention
to the simple case in which the Laplacian is $\Delta=-D^2$ with $D_\mu=\partial_\mu+A_\mu$
on a flat four dimensional manifold with Euclidean metric $\delta_{\mu\nu}$.
We also choose $A_\mu$ to be an abelian connection, to avoid the unnecessary
complications of an internal non-commuting fiber structure.
Since the two-point function is known to be transverse,
we can directly compute it by contracting it with the transverse projector and obtain only its transverse component.
There are two terms in the expansion
\begin{eqnarray}\label{2pf_RS_AA}
 \left.\frac{\delta^2\Tr K^s}{\delta A_{p;\mu}\delta A_{-p;\nu}}\right|_{\Phi=0}
 &=&
 2 \int_\theta \myincludegraphics{13}{0.3}{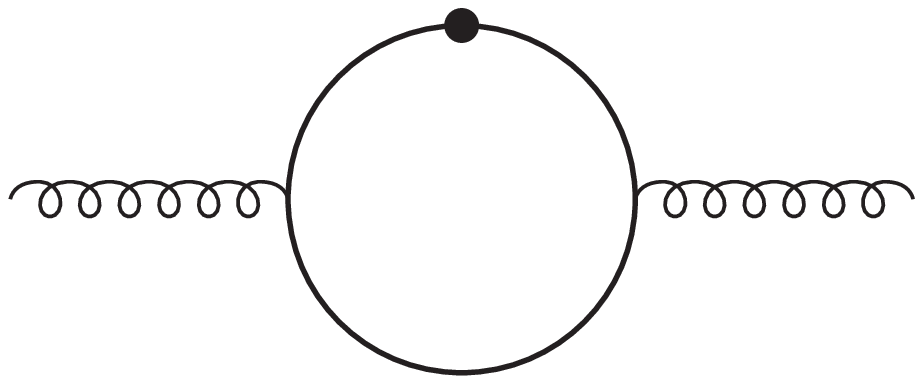} - \int_\theta \myincludegraphics{17}{0.3}{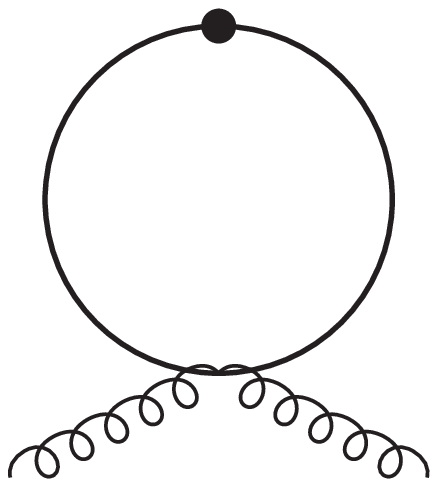}\,.
\end{eqnarray}
We explicitly included a marked point in the expansion, because the propagator in the resolvent method
does not satisfy the composition rule that is proper for the ordered product \eqref{HK_PT_5.1}. This notation should make
evident that there are three propagators in the first graph and two in the second.

The transverse part of the first diagram of \eqref{2pf_RS_AA} in momentum space is found to be
\begin{eqnarray}\label{RM_5}
 \int_{\cal C}\frac{id\theta}{2\pi}\int_{0}^{\infty}\frac{1}{2}\alpha^2 d\alpha\int_{0}^{1}d\xi\int_{q}\frac{4q^{2}}{d}e^{-\alpha(q^{2}-\theta+p^{2}\xi(1-\xi))}e^{-s\theta}\,,
\end{eqnarray}
where we used a standard relation of perturbation theory to combine the propagators
\begin{eqnarray}\label{exponentiation}
 \frac{1}{A^2B}
 =
 \int_{0}^{1}d\xi\frac{1}{(\xi A+(1-\xi)B)^{3}}=\frac{1}{2}\int_{0}^{\infty}\alpha^2 d\alpha\int_{0}^{1}d\xi e^{-\alpha(\xi A+(1-\xi)B)}\,,
\end{eqnarray}
and we shifted the momentum integration in such a way that the exponential
is a function of $q^{2}$ only. To obtain the final result one has
to perform four integrations; it proves convenient to perform the $\alpha$
integration first.
This integration has a convergence radius that depends on $\theta$ and therefore the result has to be analytically continued as anticipated.
A singularity in $\theta$ appears explicitly in
the denominator
\begin{eqnarray}\label{RM_6}
 \int_{\cal C}\frac{id\theta}{2\pi}\int_{0}^{1}d\xi\int_{q}\frac{2q^{2}\alpha^2 e^{-s\theta}}{\left[q^{2}-\theta+p^{2}\xi(1-\xi)\right]^{2}}\,.
\end{eqnarray}
Now we use the theorem of residues to perform the $\theta$ integration, where
the only pole is $\theta_{*}=q^{2}+p^{2}\xi(1-\xi)$.
Further, we integrate
over the momentum $q$ and the parameter $\xi$ to obtain
\begin{equation}\label{RM_7}
 \frac{1}{d}\int_{0}^{1}d\xi\int_{q}\frac{i}{2\pi}\,2\pi i\,{\rm Res}_{\theta_{*}}\frac{4q^{2}e^{-s\theta}}{\left[q^{2}-\theta+p^{2}\xi(1-\xi)\right]^{2}}
 =
 \frac{1}{(4\pi s)^{d/2}}2f(sp^{2})\,.
\end{equation}

The tadpole diagram of \eqref{2pf_RS_AA} has only two propagators to exponentiate
and these share the same momentum $q_\mu$, therefore
no parametric $\xi$ integration will be present.
Again, we will take into account only its transverse part, since we know that the final result
has to be transverse in the incoming momentum.
All integrations, with the exception of the parametric one, are performed in
the same order as for \eqref{RM_6} and \eqref{RM_7}, using the residue theorem at
$\theta_{*}=q^{2}$.
We obtain
\begin{eqnarray}\label{RM_8}
 \int_{\cal C}\frac{id\theta}{2\pi}\int_{0}^{\infty}\alpha d\alpha\int_{q}2e^{-\alpha(q^{2}-\theta)}e^{-s\theta}
 & = & -\int_{\cal C}\frac{id\theta}{2\pi}\int_{q}\frac{2e^{-s\theta}}{\theta-q^{2}}\nonumber \\
 & = & \int_{q}{\rm Res}_{\theta_{*}}\frac{2e^{-s\theta}}{\theta-q^{2}}\nonumber \\
 & = & \frac{2}{(4\pi s)^{d/2}}\,.
\end{eqnarray}
Needless to say, it is possible to explicitly check that the longitudinal parts of the two diagrams
of \eqref{2pf_RS_AA} cancel each other.

We finally combine the contributions from the two diagrams \eqref{RM_7} and \eqref{RM_8}
and compare with the transverse part of the Hessian \eqref{ansatz1} shown in \eqref{2pf_AA3}.
The result is
\begin{equation}
f_{\Omega}(x)=-\frac{1}{2x}\bigl[f(x)-1\bigr]\,,
\end{equation}
that agrees with \eqref{fOmega}. It is interesting to note that, within the resolvent method,
the parametric $\xi$-integration comes from the exponentiation of the propagators of \eqref{RM_5} using \eqref{exponentiation}.
This is another manifestation of formula \eqref{RM_2}, that relates the heat kernel with the standard propagator of quantum field theory.

We are not going to reproduce all the form factors with the resolvent method,
since it would not be particularly instructive. However, it is important to stress that
this method proves powerful when dealing with operators with a more complex structure
than a Laplace-like one. In particular in \cite{Gusynin_1997} the heat kernel of
a non-minimal operator is computed in its local form.
Our results show how it is
possible, with the momentum space technique, to address the issue of computing directly
the non-local form factors of a non-minimal operator.
We plan to come back to this
issue in a future publication.

\section{Conclusions}

We proposed an algorithm to compute the non-local heat kernel expansion
of a general Laplace-like operator on a Riemannian manifold, with a general connection and endomorphism term.
The method is based on a Feynman-like set of diagrammatic rules
and therefore is suited to incorporate all the strengths of the reduction
algorithms developed for quantum field theoretical perturbation theory.

The key idea of the method is that the diagrammatic rules
are entirely derived from a single object that we called Laplacian action,
that serves as a device to construct the perturbation theory of the heat kernel,
through a generating function method.
The analogy of the heat kernel with a perturbative quantum field theory
extends further: the existence of the Laplacian action
allows to interpret the heat kernel as the generating functional
of a quadratic and therefore solvable theory of a field of density weight $w=1/2$.
This field possesses a non-standard propagator and is coupled
to endomorphism, connection and metric as sources.

All the computations we did were performed in momentum space.
We used the momentum space technique to compute the expansion of the heat kernel
up to second order in a set of generalized curvatures.
Third order form factors can be derived along the same lines,
although the results are comprehensively much more complex.
The method is particularly powerful since it can be generalized easily
to any given differential operator, including differential operators that are
non-minimal and not of second order.

We introduced the resolvent method as an alternative technique,
that may be useful when dealing with more complex situations, in particular
for non-minimal operators. The resolvent method pushes the analogy
between the heat kernel and quantum field theory further. In fact
the perturbation theory that is developed using the resolvent
shares the standard propagator of a field theory.
We believe our method will prove very powerful in taming
the very complex structure of the bare form factors of the heat kernel expansion,
since it can borrow easily all the sophisticated techniques that have been developed
for the computation of perturbative quantum field theory.

It is important to stress that our momentum space technique is very versatile
and it should be evident that, with some trivial generalization, it can incorporate
the computation of heat traces involving non-trivial operator insertions easily.
As an example, one may want to compute the non-local form factors of the heat kernel trace in presence
of covariant derivative insertions \cite{Barvinsky:1991aq}.
Therefore, our method can provide a non-local generalization of the results of \cite{Decanini:2005gt}.

\section*{Acknowledgments}

We would like to thank A. Nink and R. Percacci for carefully reading the manuscript and suggestions in the phase of editing the paper.
The research of O.~Z.\ is supported by the Deutsche Forschungsgemeinschaft (DFG)
within the Emmy-Noether program (Grant SA/1975 1-1).

\appendix

\section{Useful relations}\label{appendixA}

On a $d$-dimensional manifold one can define the Euler integrand
\begin{eqnarray*}
 E & = & R_{\mu\nu\alpha\beta}R^{\mu\nu\alpha\beta}-4R_{\mu\nu}R^{\mu\nu}+R^{2}
\end{eqnarray*}
which is a total derivative in $d=4$. The Euler integrand is not a total derivative for any $d$.
However, using Bianchi identities, one can prove \cite{Avramidi_1990_2002} that for $n\ge 0$
\begin{eqnarray}
 \int d^{d}x\left[ R_{\mu\nu\alpha\beta}\square^nR^{\mu\nu\alpha\beta}-4R_{\mu\nu}\square^nR^{\mu\nu}+R\square^nR \right] = O(\mathcal{R}^3)\,; \label{Avr}
\end{eqnarray}
it follows that at order $\mathcal{R}^2$ there are only two independent form factors that can appear in \eqref{ansatz1},
since one can reduce their number by one by using \eqref{Avr} in a Taylor expansion of these.
In general, we are free to eliminate the Riemann form factor 
\begin{eqnarray*}{}
\int d^{d}x \,R_{\mu\nu\alpha\beta}f_{Riem}(s\square)R^{\mu\nu\alpha\beta}=\int d^{d}x \left\{4R_{\mu\nu}f_{Riem}(s\square)R^{\mu\nu}-Rf_{Riem}(s\square)R \right\} +O(\mathcal{R}^3)
\end{eqnarray*}
by absorbing $f_{Riem}(x)$ into the definitions of $f_{Ric}(x)$ and $f_{R}(x)$. This leads to the ansatz \eqref{ansatz1}.

In a very similar way one can change the form factor basis from the one considered in this paper $\{R_{\mu\nu},R,\U,\Omega_{\mu\nu}\}$ to the Weyl basis $\{C_{\mu\nu\alpha\beta},R,\U,\Omega_{\mu\nu}\}$,
where $C_{\mu\nu\alpha\beta}$ is the Weyl conformal tensor, the square of which is
\begin{eqnarray*}
C^{2} =  R_{\mu\nu\alpha\beta}R^{\mu\nu\alpha\beta}-\frac{4}{d-2}R_{\mu\nu}R^{\mu\nu}+\frac{2}{(d-1)(d-2)}R^{2}\,.
\end{eqnarray*}
In this case we use the following relation that is valid up to third order in the curvatures
\begin{eqnarray*}{}
 \int d^{d}x \left\{R_{\mu\nu}f_{Ric}(s\square)R^{\mu\nu}+Rf_{R}(s\square)R \right\}
 &=&
 \int d^{d}x \Bigl\{\frac{d-2}{4(d-3)}C_{\mu\nu\alpha\beta}f_{Ric}(s\square)C^{\mu\nu\alpha\beta}
 \\
 {}
 &&
 +R\Bigl[ \frac{d}{4(d-1)}f_{Ric}(s\square)+f_{R}(s\square)\Bigr] R \Bigr\}
 \,,
\end{eqnarray*}
to define the form factors:
\begin{eqnarray*}
 f_{C}(x) & = & \frac{d-2}{4(d-3)}f_{Ric}(x)\\
 f_{Rbis}(x) & = & \frac{d}{4(d-1)}f_{Ric}(x)+f_{R}(x)\,;
\end{eqnarray*}
which are those introduced in equation \eqref{HK_2.2202} of subsection \ref{Weyl_basis}.

Finally we write the form factors in the basis used in subsection \ref{BVbasis}.
The authors of \cite{Barvinsky_Vilkovisky_1987_1990} defined $\U\equiv-\mathbf{P}+\mathbf{1}R/6$
and computed the order ${\cal R}^2$ part of the trace of the heat kernel in the form
\begin{eqnarray*}{}
 \int \! d^{d}x \tr \Bigl\{\mathbf{1}R_{\mu\nu}f_{1}(s\square)R^{\mu\nu}+\mathbf{1}R f_{2}(s\square)R
 +\mathbf{P}f_{3}(s\square)R+\mathbf{P}f_{4}(s\square)\mathbf{P}+\Omega_{\mu\nu}f_{5}(s\square)\Omega_{\mu\nu} \Bigr\} \! =
 \\
 {}
 \int \! d^{d}x \tr \Bigl\{\mathbf{1}R_{\mu\nu}f_{1}(s\square)R^{\mu\nu}+\mathbf{1}R\Bigl[f_{2}(s\square)+\frac{1}{6}f_{3}(s\square)
 +\frac{1}{36}f_{4}(s\square)\Bigr]R
 \\
 {}
 +\mathbf{U}\Bigl[-f_{3}(s\square)-\frac{1}{3}f_{4}(s\square)\Bigr]R+\mathbf{U}f_{4}(s\square)\mathbf{U}
 +\Omega_{\mu\nu}f_{5}(s\square)\Omega_{\mu\nu}  \Bigr\}\,.
\end{eqnarray*}
We can match the right hand side of this expression with \eqref{ansatz1}. In this way we find the following relations between the two sets of form factors:
\begin{eqnarray*}
 f_{UR}(x) & = & -f_{3}(x)-\frac{1}{3}f_{4}(x)\\
 f_{U}(x) & = & f_{4}(x)\\
 f_{R}(x) & = & f_{2}(x)+\frac{1}{6}f_{3}(x)+\frac{1}{36}f_{4}(x)\\
 f_{Ric}(x) & = & f_{1}(x)\\
 f_{\Omega}(x) & = & f_{5}(x)\,;
\end{eqnarray*}
by inverting these we obtain \eqref{BV_basis}.

\section{Heat kernel in the coincidence limit}\label{appendixB}

As pointed out in subsection \ref{gravity_contributions} it is not possible to compute the coefficient $g_{R,0}$, that appears in \eqref{ansatz1},
by considering the momentum space diagrammatic we developed for the trace of the heat kernel.
The reason is that the momentum conservation of the $1$-point graviton function forces the incoming momentum to be zero
and therefore kills the first variation of \eqref{ansatz1} with respect to the metric.
In other words, the first variation of \eqref{ansatz1}, when evaluated in flat space,
is a total derivative and therefore cancels because of boundary conditions.
In this appendix we provide an extension of \eqref{diagrams},
that allows to compute the heat kernel and therefore has access to the coefficient $g_{R,0}$.
For simplicity, we limit ourselves to the first order in the curvatures.

We begin with providing the most general ansatz for the heat kernel, like we did for the trace of the heat kernel in \eqref{ansatz1}.
It is
\begin{eqnarray}\label{ansatz2}{}
 K^{s}(x,x)
 & = &
 \frac{1}{(4\pi s)^{d/2}}
 \sqrt{g}
 \Bigr\{
 g_0 \,\mathbf{1}
 +s \, g_{U}(s\Box) \, \U
 +s \, \mathbf{1} \, g_{R}(s\Box) \,  R
 +{\cal O}(\mathcal{R}^{2})\Bigl\} \,,
\end{eqnarray}
where two new non-local form factors $g_{U}(x)$ and $g_{R}(x)$ have been introduced.
The heat kernel of \eqref{ansatz2} has been considered in the limit of coinciding coordinates $y=x$, also called coincidence limit.
We can Taylor expand these two functions and integrate in $x$ to compare \eqref{ansatz2} with \eqref{ansatz1},
to discover that only the first term of the expansions of the non-local form factors does survive the boundary conditions.
The comparison also shows that the consistency between the two ansatzs strictly requires $g_{U}(0)=g_{U,0}$ and $g_{R}(0)=g_{R,0}$.
Our task is then to compute $g_{U}(x)$ and $g_{R}(x)$, that will uncover these two additional non-local structures,
as well as give a check for the computation of $g_{U,0}$ and a determination of $g_{R,0}$.

The diagrammatic expansion of the heat kernel is easily obtained considering \eqref{HK_PT_8}, that has been represented graphically in \eqref{HK_PT_8.1}.
Functional derivatives with respect to the fields $\Phi$ act on the cross insertions of \eqref{HK_PT_8.1} and are represented by external lines.
We are interested in the curvatures to the first order, thus the relevant diagram in the expansion is the $1$-point heat kernel diagram. It reads
\begin{eqnarray}
 \frac{\delta K^s}{\delta\Phi(x)}
 &=&
 -s \,\myincludegraphics{25.5}{0.3}{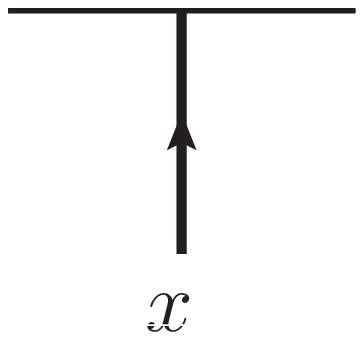}
\end{eqnarray}
and in particular we want to consider $\Phi(x)=\U(x)$ and $\Phi(x)=h_{\mu\nu}(x)$.
The case in which $\Phi(x)=A_\mu(x)$ can be shown to give a zero result at the coincidence limit.

We first compute the diagram with an external endomorphism $\U$ line in momentum space, that gives
\begin{eqnarray}
 \frac{\delta K^s}{\delta\U_p}
 &=& -s\, \myincludegraphics{25}{0.3}{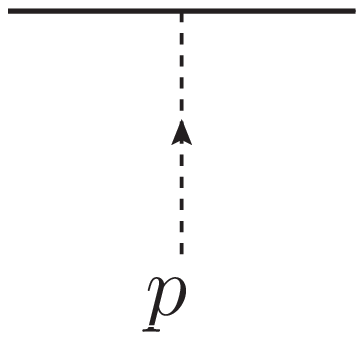}
 \nonumber
 \\
 &=&
 -s \int_q \int_0^1 d\xi e^{-s \xi q^2}e^{-s (1-\xi)(q+p)^2}\,.
\end{eqnarray}
It is trivial to manipulate the integral in the same fashion of those of section \ref{section2}, to get
\begin{eqnarray}
 \frac{\delta K^s}{\delta\U_p}
 &=&
 -\frac{s}{(4\pi s)^{d/2}}f(sp^2)\,.
\end{eqnarray}
Comparing our computation with the first variation of \eqref{ansatz2}, one obtains
\begin{eqnarray}
 g_U(x)
 &=&
 - f(x)\,,
\end{eqnarray}
that has the correct limit $g_U(0)=-f(0)=-1$.
Interestingly, also this form factor can be seen as the resummation of infinitely many terms of the local expansion of the heat kernel \cite{Avramidi_1990_2002}.
In particular, it contains all the information about the terms of the form $\Box^n \U$, that are suppressed in our original ansatz by the boundary conditions for any $n>0$.
As a further check, it is easy to see that the first order in the expansion gives the term
$$
 \frac{s^2}{(4\pi s)^{d/2}} \frac{1}{6}\Box\U\,,
$$
that agrees with the results of the local expansion \cite{Vassilevich_2003}.

We now proceed by computing $g_{R}(x)$, for which the $1$-point function at $\Phi(x)=h_{\mu\nu}(x)$ is needed.
We obtain
\begin{eqnarray}
 \frac{\delta K^s}{\delta h_{p;\mu\nu}}
 &=& -s\,\myincludegraphics{25}{0.3}{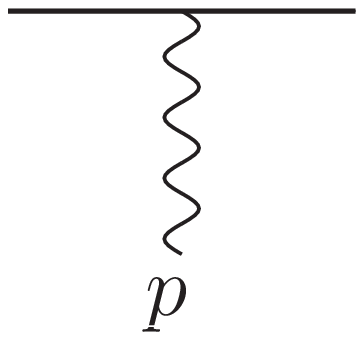}
 \nonumber
 \\
 &=&
 \frac{1}{(4\pi s)^{d/2}}
 \Bigl[
 \Bigl(\frac{1}{2}+\frac{1}{4}sp^2\Bigr)f(sp^2)P_{T}^{\mu\nu}
 +\frac{1}{2}P_{L}^{\mu\nu}
 \Bigr]
 \,,
\end{eqnarray}
that has to be compared with the same quantity derived from \eqref{ansatz2}.
As expected, the longitudinal component of this result does not depend on the incoming momentum, since
its contribution arises solely from the first (volume) term of \eqref{ansatz2}. The complete comparison
allows to determine $g_R(x)$ as
\begin{eqnarray}
 g_R(x)
 &=&
 -\frac{1}{2x}+\frac{1}{4x} (x+2) f(x)\,,
\end{eqnarray}
and, in particular, we find
\begin{eqnarray}
 g_{R,0}=g_R(0)=\frac{1}{6}\,.
\end{eqnarray}
As for the previous case, the form factor $g_R(x)$ can be understood as the resummation of infinitely many terms of the local expansion.
In this case it resums all the possible terms of the form $\Box^n R$. We can therefore check, as an example,
the first term of the expansion that contributes to the local heat kernel as
$$
 \frac{s^2}{(4\pi s)^{d/2}} \frac{1}{30}\Box R\,,
$$
that agrees with the results of the local algorithms \cite{Vassilevich_2003}.

It is interesting to note that the two form factors computed in this appendix, $g_{U}(x)$ and $g_{R}(x)$,
share the same basic form factor \eqref{HK_2.3} with the system \eqref{HK_2.22}.
This is a general feature of the expansion of the heat kernel.
Namely, the heat kernel at $n$-th order in the curvatures
will share the same basic form factors of the trace of the heat kernel at $(n+1)$-th order
as shown in \cite{Barvinsky:1991aq}.

\newpage


\end{document}